\documentclass[preprint,12pt,sort]{elsarticle}

\usepackage{amssymb}
\usepackage{amsmath}
\usepackage[colorlinks=true,
            linkcolor=cyan,
            citecolor=cyan,
            urlcolor=blue]{hyperref}

\usepackage{algorithm}
\usepackage{algorithmic}
\usepackage{epsfig}
\usepackage{subfigure}
\usepackage{multirow}
\usepackage{float}
\usepackage{amsmath}
\usepackage{longtable}
\usepackage{booktabs}
\usepackage{color}
\usepackage{fancyhdr}
\usepackage{epstopdf}
\usepackage{ulem}
\usepackage{bm}
\usepackage{amsopn}
\usepackage{amssymb,amsmath,color,times}
\usepackage{lineno}
\usepackage{times}
\usepackage{xcolor}
\usepackage{amsfonts}
\usepackage{soul}
\usepackage{algorithm}
\usepackage{multirow}
\usepackage{graphicx}
\usepackage{subfigure}
\usepackage{epstopdf}
\usepackage{multicol}
\usepackage{algorithmic}
\usepackage[utf8]{inputenc}
\usepackage[small]{caption}
\usepackage{array}
\usepackage{verbatim}
\usepackage{subcaption}
\usepackage{geometry}

\fancypagestyle{titlepage}{
    \fancyhf{} 
}

\begin{document}

    
    
    


    


\begin{frontmatter}

\title{DTD-VAE: Disentangled Temporal Dependencies VAE for Credit Risk Prediction} 


\author[affi_1]{Xiaobo Guo}

\author[affi_1]{Lu-an Dong}

\author[affi_3]{Yanbo Wang}

\author[affi_4]{Peng Zhang}

\author[affi_2]{Cai Zhi}

\author[affi_2]{Youru Li}
\ead{liyouru@bjut.edu.cn}

\affiliation[affi_1]{organization={Data Management Department, China Minsheng Bank},
            postcode={100010},
            state={Beijing},
            country={China}}

\affiliation[affi_2]{organization={College of Computer Science, Beijing University of Technology},
            postcode={100044},
            state={Beijing},
            country={China}}

\affiliation[affi_3]{organization={Data Intelligence Division, Longying Zhida (Beijing) Technology},
            postcode={100020},
            state={Beijing},
            country={China}}

\affiliation[affi_4]{organization={Cyberspace Institute of Advanced Technology, Guangzhou University},
            postcode={510006},
            state={Guangzhou},
            country={China}}

\begin{abstract}
Evaluating customer creditworthiness is crucial for retail banking operations, as it impacts marketing strategies, customer relationship management, and credit risk control. Traditional methods often struggle to capture complex temporal dependencies and extract pertinent information from customer data, crucial for accurate risk assessment. Specifically, they fail to differentiate between temporal patterns indicative of credit risk and those reflecting general customer behavior or preferences, leading to suboptimal risk predictions. In this study, we introduce the Disentangled Temporal Dependencies Variational Autoencoder (DTD-VAE), an advancement over conventional VAE, designed to disentangle temporal dependencies and distinguish credit risk-related features from past customer preferences. The feature inference module of the DTD-VAE incorporates an autoregressive temporal dependency learning mechanism that adeptly captures the temporal dependencies among latent variables, enriching the model’s comprehension of the inherent data structure. Furthermore, the feature generative module utilizes an element-wise gating mechanism that assigns independent weights to each dimension of the expert models, enabling a finer-grained disentanglement of latent variables, particularly those relevant to credit risk prediction. Extensive experiments on six real-world datasets demonstrate that the proposed framework consistently outperforms existing methods, achieving performance gains of 3.2\%-4.86\% in ROC-AUC and 6.41\%-9.71\% in Accuracy Ratio. 
\end{abstract}


\end{frontmatter}

\section{Introduction}\label{sec:intro}
\subsection{Background}
\begin{figure}
\centering
\begin{minipage}[t]{0.45\textwidth}
    \centering
    \includegraphics[width=0.7\linewidth]{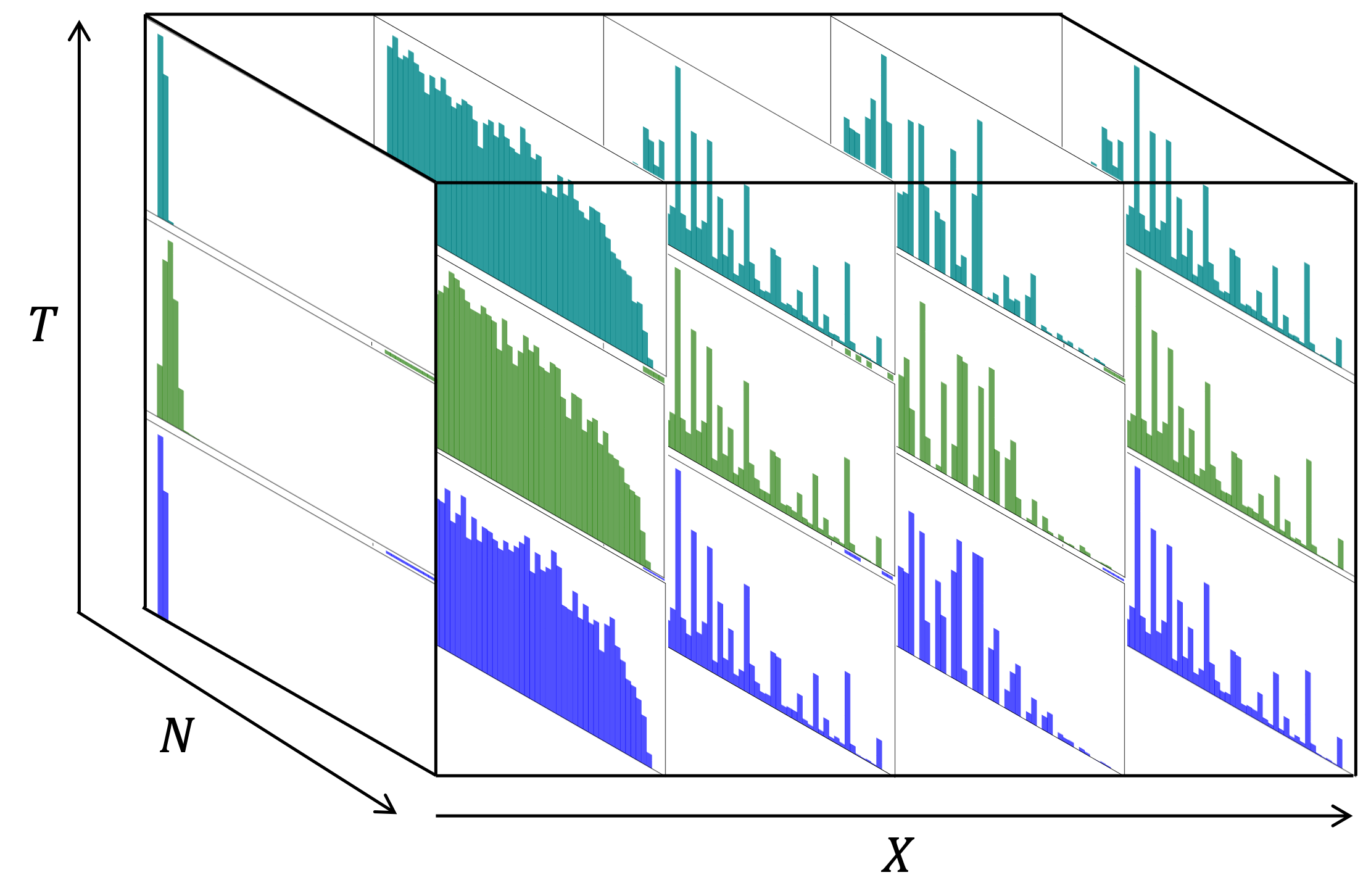}
    \caption{\footnotesize Credit data often exhibits non-stationarity, with distributions changing over time. During an economic recession, declining repayment ability can lead to increased default rates. The model must identify these abrupt changes and dynamically adjust the risk predictions. We present relevant features from the Lending Club dataset, where \(X\) denotes features, \(T\) represents temporal dependencies, and \(N\) indicates sample numbers.}
    \label{fig:_temporal_depedency_}
\end{minipage}
\hfill
\begin{minipage}[t]{0.48\textwidth}
    \centering
    \includegraphics[width=\linewidth]{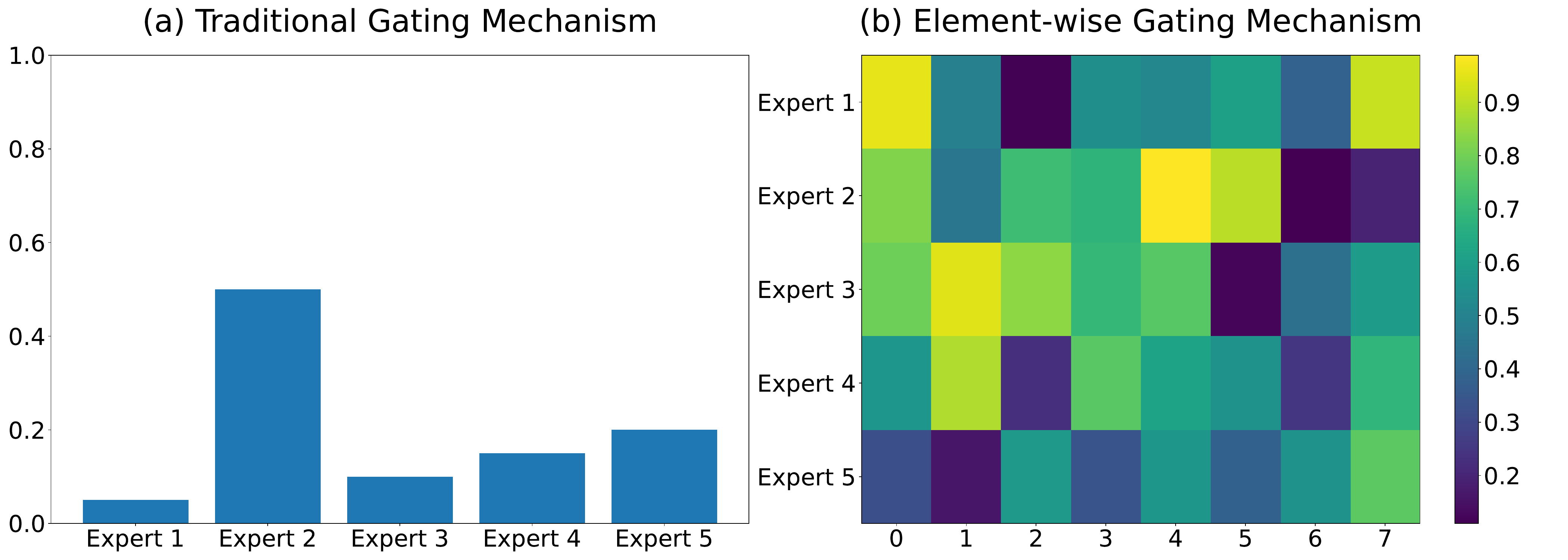}
    \caption{\footnotesize (a) Traditional Mixture of Experts (MoEs) are combined using a gating network to select the most appropriate expert for each input. (b) Our Element-wise Gating (EG) method is introduced for finer control over the selection process, improving adaptability and efficiency.}
    \label{fig:_moe_eg_weights_}
\end{minipage}
\end{figure}

In the dynamic realm of financial services, credit risk prediction is critical in risk control and management for consumer finance companies, small loan institutions, banks, and other entities~\citep{genovesi2024standardizing}. Effective prediction can help manage and mitigate loan defaults and delinquencies, keeping bad debt levels low ~\citep{cheng2020contagious,DBLP:journals/tist/TangWZJ24,lan2025sparse}. For example, by effectively predicting credit risk, financial institutions have significantly reduced losses in the multi-billion-dollar credit loan industry, enhancing their risk management capabilities~\citep{tan2018deep,zhao2023fintech,DBLP:journals/tnn/LiZZCJ25}. Similarly, in retail banking, customer risk assessment for loan default prediction is a pivotal task that requires a profound understanding of the complex and temporal nature of customer data. In previous practice, financial institutions have employed a variety of models to assess credit risk, ranging from simple loan default prediction techniques to more sophisticated statistical analyses. 
To enhance the prediction in customer risk assessment, it is essential to effectively account for the complexity and temporal dependencies within data, including both time-related and simultaneously extracted task-relevant data ~\citep{tao2018multiple}. Credit data often exhibit non-stationarity due to factors like economic cycles and policy changes, requiring models that can dynamically adapt, as depicted in Fig.~\ref{fig:_temporal_depedency_}. In contrast, marketing data are influenced by seasonal trends and promotions but generally do not experience the same frequent and drastic changes~\citep{chen2023financial,ye2024frequency}. However, customer data usually exhibit multidimensionality, featuring complex nonlinear relationships and significant temporal dependencies, which are aspects that traditional linear models find difficult to capture. In fact, it will provide deeper insights into customers' credit status and generate reliable predictions by modeling these structures and dependencies. Thus, considering data complexity, temporal dependencies, and task-relevant information, the predictions can be significantly improved. In this way, although the existing method has reshaped the evaluation of credit risk, it still faces the main challenges outlined in the following aspects:

\textbf{CH1: How can we disentangle the data complexity and temporal dependencies for loan default risk prediction?} Traditional machine learning algorithms struggle with loan default risk prediction because of data complexity and temporal dependencies. High-dimensional, nonlinear data complicate feature capture, impacting the accuracy of methods such as Random Forests and LightGBM. In addition, these algorithms fail to account for temporal dependencies in time-series data (e.g., historical credit scores), leading to inaccurate predictions. By contrast, deep learning offers a more effective solution. Deep latent variable models trained with amortized variational inference have advanced representation learning on high-dimensional datasets~\citep{kingma2013auto}. Recent studies have enhanced RNNs by introducing stochastic latent variables within the variational autoencoder (VAE) framework~\citep{kingma2014stochastic,fraccaro2016sequential}. The VAE enables end-to-end training by using neural networks to parameterize both the posterior and generative models. However, these models often rely on simple latent distributions, such as Gaussian, and assume independence among latent variables, which reduces their flexibility in modeling complex and temporal patterns~\citep{he2018variational,li2023distvae}, ultimately impairing their ability to fit the data accurately.

\textbf{CH2: How can task-relevant information be effectively extracted for loan default risk prediction?} Similar to \textbf{CH1}, traditional machine learning algorithms also face challenges in extracting task-relevant information for loan default risk prediction. These algorithms struggle with feature extraction and representation by assuming linear or simple tree-like relationships that fail to capture complex interactions~\citep{tan2018deep}. For instance, Random Forest uses simple feature splits and often misses intricate dependencies despite ensemble aggregation. Furthermore, these models lack effective mechanisms for modeling informative latent variables. Deep generative models, such as VDFDP~\citep{huang2024diagnosis}, MVOFS-OVAE~\citep{muthukumaran2023feature}, and recent variants~\citep{tan2024dptvae}, have been applied to learn latent representations. Although these approaches capture latent variable traits, they struggle to identify and utilize critical task-specific information. Our approach improves task performance by integrating an element-wise gating experts mechanism, replacing Multilayer Perceptrons (MLPs) decoders with multiple specialized subnetworks, and a gating network to dynamically select the optimal expert for task-specific information extraction, as shown in Fig. ~\ref{fig:_moe_eg_weights_}.

\subsection{Research objectives} 
To this end, this paper proposes the DTD-VAE, a novel principled variational autoencoder framework for loan default risk prediction, as illustrated in Fig. \ref{fig:_ds_vae_framework_}. \textbf{To cope with CH1}, the DTD-VAE's feature inference module incorporates an Autoregressive Temporal Dependency (ATD) learning mechanism to capture temporal dependencies between latent variables, thus enhancing the description of data's intrinsic structure. It simultaneously captures temporal dependencies among latent variables and evolving trends of latent variables over time, thereby providing more accurate predictions. \textbf{In response to CH2}, the feature generative module integrates an Element-wise Gating (EG) mechanism, which assigns independent weights to each dimension of the expert models, enabling finer-grained disentanglement of latent variables. This enhances the quality and diversity of the generated samples, thereby improving the interpretability and controllability of the model. By integrating these advanced modules, our DTD-VAE framework enhances the accuracy and reliability of loan default risk predictions, offering a robust solution for customer risk assessments in various financial contexts.

\subsection{Contributions}
In summary, this work represents a paradigm shift from traditional, static, and unidimensional credit risk assessment to an end-to-end generative modeling approach that jointly captures dynamic latent structures and interpretable feature representations. The main contributions are as follows:
\begin{itemize}
    \item A novel probabilistic generative framework DTD-VAE is proposed to disentangle temporal dependencies and task-specific customer preferences, enhancing robustness and generalization in credit risk prediction.

    \item As a simple yet effective mechanism, ATD is designed to model temporal dependencies and response variability, providing a general framework that enhances and complements existing temporal modeling approaches.
    
    \item An EG mechanism is incorporated to enable fine-grained disentanglement of latent variables through dimension-wise weighting, improving generative expressiveness and risk sensitivity.

    \item Extensive experiments on six real-world datasets validate the consistent superiority of DTD-VAE over state-of-the-art methods in financial risk prediction.
\end{itemize}

The remainder of this paper is organized as follows: Section~\ref{sec:related_work} reviews the related work. Section~\ref{sec:prelim} provides the necessary background and outlines the research question. Section~\ref{sec:method} presents the details of the proposed method. Section~\ref{sec:exper} describes the experimental methodology and analyzes the results. Finally, Section~\ref{sec:conclusion} concludes the paper and suggests directions for future work.

\section{Related Work}\label{sec:related_work}

\subsection{Financial Risk Prediction Modeling}
Financial risk prediction is vital for managing lending portfolios, informing decisions, and maintaining financial stability~\citep{locurcio2021credit,jan2022data}. Traditional methods include credit scoring models, financial ratios, and statistical techniques, all of which have distinct advantages and limitations. Credit scoring models that assess risk based on credit history and other factors are critical for setting credit terms and interest rates~\citep{pang2021borrowers}. These models can be industry-specific (e.g., automotive and mortgage) or customized to fit lenders' criteria and risk preferences~\citep{ashofteh2021conservative}. Recently, machine learning techniques such as Logistic Regression, Decision Trees, Random Forests, and Support Vector Machines have shown promise in credit risk assessment. However, most methods lack temporal features, and studies suggest that XGBoost and LightGBM often outperform deep learning approaches~\citep{xu2021loan}.  Despite the benefits of incorporating additional structural information, predefined dependency structures may limit performance~\citep{he2018variational}.

\subsection{Temporal Dependencies Disentangling}
Deep learning has facilitated the development of sequential probabilistic models for high-dimensional data~\citep{girin2020dynamical, fraccaro2016sequential}.Techniques such as memory units and stochastic latent variables improve the capture of these complex dynamics. However, current models often fail to fully capture all temporal dependencies. The VAE is popular for modeling complex distributions by approximating intractable posteriors. Chung et al.~\citep{chung2015recurrent} integrated the VAE into LSTM and explored stochastic units in RNNs. Fraccaro et al.~\citep{fraccaro2016sequential} proposed stochastic RNNs with autoregressive structures, and Xu and Chen~\citep{xu2021deep} applied similar models to financial data. FHVAE~\citep{hsu2017unsupervised} introduced a factorized hierarchical model for speech data; however, it did not leverage sequential prior knowledge. Our approach aims to fully disentangle latent variables into temporal and informative components and characterize temporal dependencies to explain latent feature interactions.

\subsection{Gating Mechanism-incorporated Approaches}
Mixture of Experts (MoEs) is a classical ensemble learning technique that efficiently scales up the model capacity ~\citep{jacobs1991adaptive}. They have been widely adopted in healthcare, finance, and in pattern recognition. MoE-based layers enhance the computational efficiency of large neural networks and improve parameter sharing~\citep{shazeer2017outrageously}. VAE~\citep{kingma2013auto} jointly train generative models and inference networks. The multimodal VAE models the joint posterior as a product of experts over marginal posteriors, enabling cross-modal generation~\citep{shazeer2017outrageously}. Shi et al.~\citep{shi2019variational} factorized the joint variational posterior using MoEs. Aljundi et al.~\citep{aljundi2017expert} proposed an Expert Gate model for lifelong learning by selecting the most relevant autoencoder based on reconstruction errors. Our work integrates an EG mechanism with VAE to address generative representation learning challenges in financial risk prediction, with the aim of improving performance, scalability, and user experience.

\section{Preliminaries}\label{sec:prelim}

\subsection{Problem Statement}
Dataset $\mathcal{D} = \{ (x_i, y_i) \}_{i=1}^N$ consists of $N$ historical loan records, where each $x_i \in \mathcal{X}$ is a feature vector describing the $i$-th application, and $y_i \in \mathcal{Y} = \{0, 1\}$ is the binary default label. The feature vector $x_i = [x_{i1}, \dots, x_{im}]$ contains $m$ attributes characterizing the applicant and the loan. The goal is to learn a model $f: \mathcal{X} \rightarrow [0, 1]$ that maps $x_i$ to a default probability $\hat{y}_i = f(x_i)$, trained on $\mathcal{D}$ to minimize the prediction error, which is typically evaluated by the ROC-AUC.

\subsection{Variational Autoencoder}
The VAE is an unsupervised deep generative model composed of an encoder and decoder. The encoder (inference model) maps input data $x$ to a latent representation $z$ via a variational posterior $q_\phi(z|x)$, while the decoder (generative model) reconstructs $x$ from $z$ through the conditional distribution $p_\theta(x|z)$. The generative process involves i) sampling latent variables $z \sim p(z) = \mathcal{N}(0, I)$ and ii) generating data $x \sim p_\theta(x|z)$. The model parameters $\theta$ and $\phi$ are typically learned using neural networks.

The VAE objective is given by
\begin{equation}
\label{eq:l_vae}
\mathcal{L}_{\text{VAE}}(\theta, \phi) = \mathbb{E}_{q_\phi(z|x)}[ - \log p_\theta(x|z) ] + \mathbb{D}_\text{KL}[q_\phi(z|x) \| p(z)],
\end{equation}
which balances the reconstruction accuracy and regularization via the KL divergence between $q_\phi(z|x)$ and the prior $p(z)$. Multilayer perceptrons (MLPs) are commonly used to parameterize both the inference and reconstruction processes. In this study, $\phi$ and $\theta$ denote the parameters of the inference and generative networks, respectively.

\section{Methodology}\label{sec:method}

\subsection{Overview}
\begin{figure*}
\centering
\includegraphics[width=0.7\textwidth]{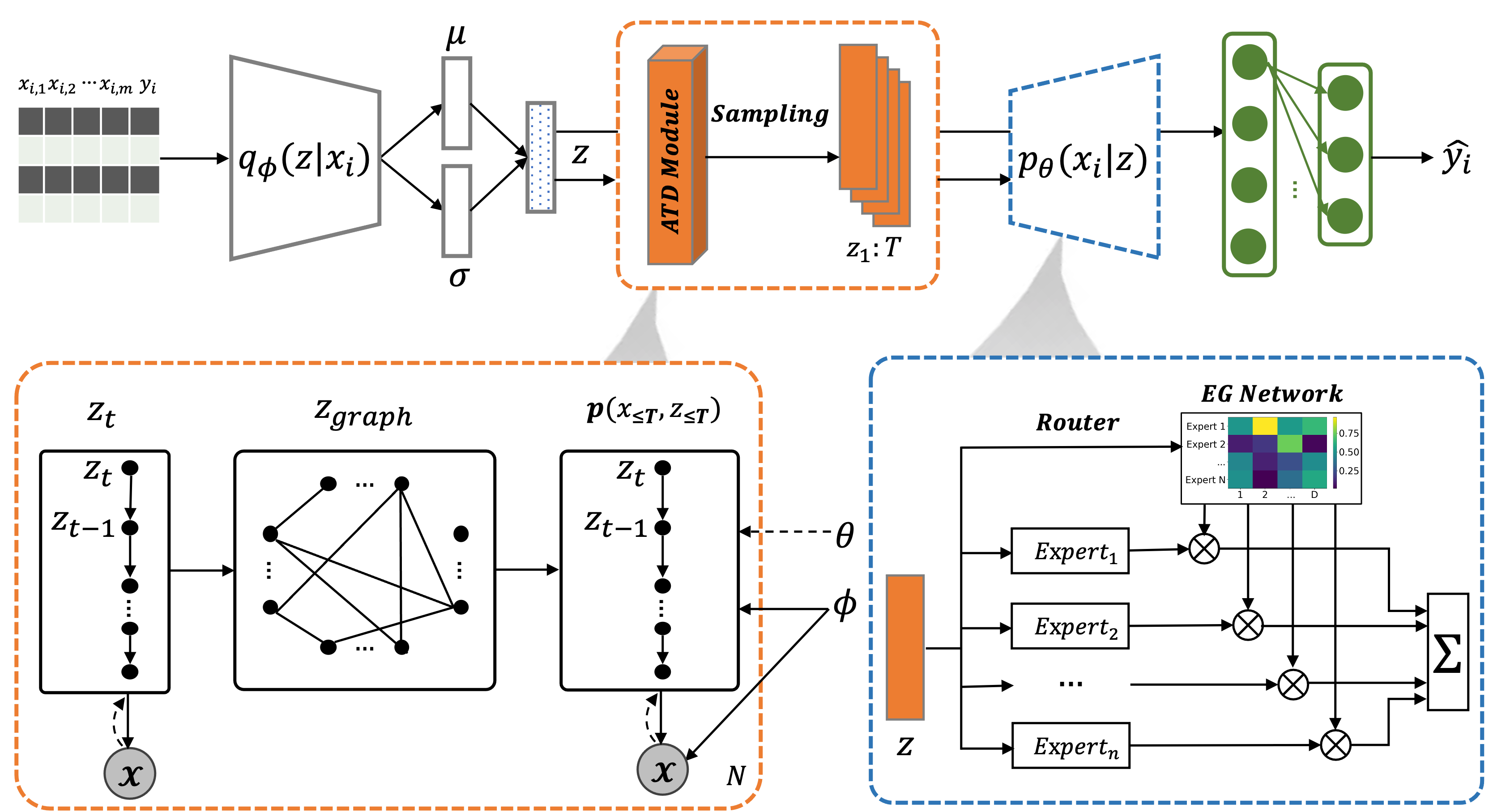}
\caption{\footnotesize The ATD Module (yellow part) captures the temporal dependencies, enhancing the model's understanding of the intrinsic structure of the data. The EG Module (blue part) assigns independent weights to different dimensions of expert models, facilitating a finer-grained disentanglement of latent variables, particularly those relevant to credit risk prediction. The Prediction Module (green part) uses an integration and joint prediction for end-to-end classification.}
\label{fig:_ds_vae_framework_}
\end{figure*}

As shown in Fig.~\ref{fig:_ds_vae_framework_}, we designed an end-to-end DTD-VAE model for financial risk prediction, consisting of three main components: \textbf{i)} Temporal Dependencies Learning via ATD, which transforms the temporal prior distribution into a graph model using an adjacency matrix to capture complex relationships and temporal dependencies among latent variables; \textbf{ii)} Task-Relevant Information Learning with EG, which assigns independent weights to each dimension of expert models to disentangle pertinent latent variables and improve the accuracy and interpretability of credit risk predictions; and \textbf{iii)} Integration and Joint Prediction, which involves jointing the final loss function from preceding components using simple MLPs to generate risk predictions. This end-to-end framework enables joint optimization of all components through unified variational inference, promoting coherent learning dynamics and enhancing both model performance and robustness.

\subsection{Temporal Dependencies Learning via ATD}
As discussed in Section~\ref{sec:intro}, ATD can be integrated into various stages of the forward process, including inference, latent model, and output space. For instance, Inverse Autoregressive Flow (IAF)~\citep{kingma2016improved} models such dependencies during inference by transforming a simple prior into a structured posterior. Hierarchical latent structures~\citep{sonderby2016ladder} further enhance model expressiveness by introducing dependencies among latent variables, which serve as learned empirical priors. In loan default prediction, where borrower conditions evolve over time, we propose an autoregressive mechanism to capture temporal dependencies among latent variables without predefined structures, thereby enabling the model to dynamically adapt to complex temporal patterns in the data.

\subsubsection{Latent Variable Sampling}
In the VAE, the sampling process of the latent variables is achieved using a reparameterization trick. Specifically, given the mean \(\mu\) and variance \(\sigma^2\) from the encoder, the latent variable \( z \) is sampled using the reparameterization trick:
\begin{equation}
z = \mu + \sigma \cdot \epsilon,
\end{equation}
where \(\epsilon \sim \mathcal{N}(0, 1)\). This ensures that \( z \) is distributed as \( \mathcal{N}(\mu, \sigma^2) \). For latent variables, we can extend this to a sequence of latent variables \( z_1, z_2, \dots, z_T \):
\begin{equation}
z_t = \mu_t + \sigma_t \cdot \epsilon_t,
\end{equation}
where \(\mu_t\) and \(\sigma_t\) are functions of the previous latent variables \( z_{<t} \) and the current input \( x_t \). This formulation ensures that \( z_t \) captures both the local information of \( x_t \) and temporal dependencies from previous time steps.

\subsubsection{Empirical Priors Informed by Latent Temporal Dependencies} Latent temporal dependencies refer to the interdependencies among latent variables across different time steps in a time series. In many real-world applications, such as financial time series analysis, the latent variables \( z_1, z_2, \dots, z_T \) are typically not independent but follow an autoregressive process. To capture these dependencies, we introduced a temporal prior distribution. Suppose that the latent variable \( z_t \) at time \( t \) depends on the state of the latent variable at the previous time step \( t-1 \) and introduces a noise term \(\epsilon_t\):
\begin{equation}
z_t = f(z_{t-1}) + \epsilon_t,
\end{equation}
where \( f(z_{t-1}) \) is a function representing the latent variable of the current time step that depends on the latent variable of the previous time step. Suppose \( f(z_{t-1}) \) is a linear function:
\begin{equation}
f(z_{t-1}) = T z_{t-1} + b,
\end{equation}
where \( T \) is the state transition matrix, and \( b \) is the bias term. The noise term \(\epsilon_t\) is typically assumed to be Gaussian \(\mathcal{N}(0, \sigma^2)\). The Ladder VAE ~\citep{webb2018faithful,sonderby2016ladder} introduces a hierarchical structure for latent variables but does not model temporal dependencies. In contrast, our approach explicitly models these dependencies, making it more effective for time-series data. For example, in financial data, \( z_t \) represents a latent variable at time \( t \). While the Ladder VAE conditions \( z_t \) on higher-layer variables, our approach conditions \( z_t \) on \( z_{t-1} \) to capture the dynamics between consecutive time steps, which is crucial for predicting loan defaults.

\subsubsection{ Latent Temporal Graph Construction}
To capture the evolving interdependencies among latent variables, we constructed a latent temporal graph that encodes dynamic relationships through a learnable topology. Given a sequence of latent states $ z_1, z_2, \dots, z_T $, we model their interdependencies via a learnable adjacency matrix $ A \in \mathbb{R}^{d_z \times d_z} $, where each element $ A_{ij} $ represents the strength of influence from latent variable $ z_j $ to $ z_i $. Rather than predefined or binary connections, $ \mathbf{A} $ is initialized as a parameter matrix and optimized during training to capture latent relational patterns. The graph-enhanced representation $ z_t^{\text{graph}} $ is then computed as:
\begin{equation}
    z_t^{\text{graph}} = A z_t,
\end{equation}
This performs a linear transformation of the latent state through the learned relational structure, effectively simulating risk propagation in the latent space while preserving the temporal dynamics.

\subsubsection{Encoding of Entire Latent Temporal Dependencies}
To better capture the temporal dynamics of the entire time series, we employed a Gated Recurrent Unit (GRU) after transforming the data using a graph model~\citep{cho2014learning}. This choice is justified by the GRU's ability to handle long-term dependencies, which is crucial for capturing the temporal evolution of credit risk. Specifically, we imposed temporal dependencies on consecutive subsequences, such as \( p(z_t, z_{t-1}) = p(z_t | z_{t-1}) p(z_{t-1}) \), and used the GRU's hidden states to capture long-term dependencies across other subsequences.
The hidden state \( h_t \) is calculated as:

\begin{equation}
  \begin{aligned}
    h_t &= \text{GRU}(h_{t-1}, x_t), \\
    r_t &= f(W_r[h_{t-1}; x_t]), \\
    u_t &= f(W[h_{t-1}; x_t]), \\
    \tilde{h}_t &= \tanh(W_{\tilde{h}}[r_t \odot h_{t-1}; x_t]), \\
    h_t &= (1 - u_t) \odot h_{t-1} + u_t \odot \tilde{h}_t.
  \end{aligned}
\end{equation}
where \( r_t \) and \( u_t \) are the reset and update gate vectors, respectively, \( \tilde{h}_t \) is the candidate activation vector, and \( h_t \) is obtained by balancing short-term memory \( h_{t-1} \) and long-term memory \( \tilde{h}_t \).

In the temporal dependency learning process, we aim to capture complex dependencies in time-series data. The generative model factorizes the joint distribution of the observed data \( x_{\leq T} \) and latent variables \( z_{\leq T} \) to reflect temporal dependencies and preserve local patterns. The model is factorized as:
\begin{equation}
 p(x_{\leq T}, z_{\leq T}) = \prod_{t=1}^T p(x_t | z_{\leq t}, x_{<t}) p(z_t | x_{<t}, z_{t-1}), 
\end{equation}
where \( p(x_t | z_{\leq t}, x_{<t}) \) is the probability of observing \( x_t \) given all previous latent variables \( z_{\leq t} \) and observed data \( x_{<t} \), and \( p(z_t | x_{<t}, z_{t-1}) \) is the probability of the latent variable \( z_t \) given the previously observed data \( x_{<t} \) and latent variable \( z_{t-1} \).

To capture the full temporal dependencies of the latent variables, we define the distributions \( p(z_t | x_{<t}, z_{t-1}) \) and \( p(x_t | z_{\leq t}, x_{<t}) \). The distribution \( p(z_t | x_{<t}, z_{t-1}) \) is modeled using a conditional Gaussian distribution \( \mathcal{N}(z_t | r_t(x_{<t}, z_{t-1}), u_t(x_{<t}, z_{t-1})) \), where \( r_t \) and \( u_t \) are functions that are parameterized by GRUs. Similarly, \( p(x_t | z_{\leq t}, x_{<t}) \) is modeled using a conditional Gaussian distribution \( \mathcal{N}(x_t | r_t'(z_{\leq t}, x_{<t}), u_t'(z_{\leq t}, x_{<t})) \), where \( r_t' \) and \( u_t' \) are also functions parameterized by GRUs. The GRU recursively calculates the hidden states \( h_t = \text{GRU}(h_{t-1}, x_t) \), where \( h_t \) is the hidden state at time \( t \), and captures the temporal dependencies. The final formal definition is:
\begin{align}
p(x_{\leq T}, z_{\leq T}) &= \prod_{t=1}^T \mathcal{N}(x_t | r_t'(z_{\leq t}, x_{<t}), u_t'(z_{\leq t}, x_{<t})) \nonumber \\ 
&\times \mathcal{N}(z_t | r_t(x_{<t}, z_{t-1}), u_t(x_{<t}, z_{t-1})).
\end{align}
This formulation ensures that the local patterns of \( x_t \) are preserved, capturing the temporal dependencies across all previous time steps.

\subsection{Task-Relevant Information Learning with EG}
The gating mechanism aims to disentangle representations by maximizing the inter-expert discrepancy across dimensions~\citep{cai2024survey,guo2024multi}. However, existing approaches like MMoE, PLE, and PEPNet have limitations: MMoE and PLE assign uniform weights across dimensions within each expert, causing mutual constraints, whereas PEPNet scales dimensions individually but lacks explicit disentanglement. To address this, we introduce an EG mechanism in the DTD-VAE decoder, which assigns independent weights per dimension per expert, enabling fine-grained disentangled representation learning. This enhances sample diversity, model interpretability, and controllability.

\subsubsection{Expert Networks Output}
Suppose that we have \( M \) expert models, each producing a \( d \)-dimensional vector \( h_m \), where \( d \) is the latent space dimension. The output of the \( m \)-th expert model can be represented as
\begin{equation}
h_m = [h_{m1}, h_{m2}, \dots, h_{md}],
\end{equation}
where \( h_{mi} \) is the \( i \)th dimension of the output vector from the \( m \)th expert model. Each expert captures distinct data aspects (e.g., features and attributes), enabling a more comprehensive input representation.

\subsubsection{Element-wise Weight Assignment}
Each dimension of the expert model was assigned an independent weight. Let \( w_m \) be the weight vector of the \( m \)-th expert model, where \( w_{mi} \) is the weight of the \( i \)-th dimension of the \( m \)-th expert model. The weight vector for the \( m \)-th expert model is:
\begin{equation}
w_m = [w_{m1}, w_{m2}, \dots, w_{md}].
\end{equation}
To ensure that the sum of the weights for each dimension is 1, we performed a softmax operation on the weight vector. Let \( a_{mi} \) be the original weight of the \( i \)-th dimension of the \( m \)-th expert model. The softmax operation is defined as follows:
\begin{equation}
w_{mi} = \frac{\exp(a_{mi})}{\sum_{j=1}^d \exp(a_{mj})}.
\end{equation}
This normalization ensures that the weights are properly scaled and summed to one across the dimensions of each model.

\subsubsection{Combined Output}
The output of each expert model is multiplied by its corresponding weight, and the weighted outputs of all expert models are summed to obtain the final disentangled representation \( z \). The final output \( z_i \) for the \( i \)th dimension is computed as follow:
\begin{equation}
z_i = \sum_{m=1}^M w_{mi} h_{mi}.
\end{equation}
This can be represented in matrix form as follows:
\begin{equation}
z = \sum_{m=1}^M w_m \odot h_m,
\end{equation}
where \( \odot \) denotes element-wise multiplication. This structured approach ensures that the model dynamically adjusts the contribution of each expert model to the final representation, thereby enhancing the flexibility and interpretability of the DTD-VAE. By combining the outputs of multiple expert models, the model can effectively balance the contributions of different features, leading to a more robust and interpretable representation. 

\subsection{Integration and Joint Prediction Module}
To predict credit risk, we employ a classifier \( f \) that maps the disentangled representation \( z \) to the credit risk score \( y \). The model \( f \) can be a simple MLP or any other suitable architecture. Formally, given \( f \), the input is \( z \) and output is \( y \):
\begin{equation}
y = f(z)
\end{equation}
The joint loss function in the proposed model was designed to balance the reconstruction error, KL divergence, and classification error. The final loss function is defined as:

\begin{equation}
\label{eq:loss_func}
\begin{aligned}
\mathcal{L}(\theta, \phi, \psi) &= \mathcal{L}_{\text{MSE}} + \mathcal{L}_{\text{KLD}} + \mathcal{L}_{\text{BCE}}, \\
\mathcal{L}_{\text{MSE}} &= \sum_{i=1}^{N} (x_i - \hat{x}_i)^2, \\
\mathcal{L}_{\text{KLD}} &= -0.5 \sum_{i=1}^{N} \left( 1 + \log(\sigma_i^2) - \mu_i^2 - \sigma_i^2 \right), \\
\mathcal{L}_{\text{BCE}} &= -\sum_{i=1}^{N} \left( y_i \log(\hat{y}_i) + (1 - y_i) \log(1 - \hat{y}_i) \right).
\end{aligned}
\end{equation}
where \(\mathcal{L_{\text{MSE}}}\) is the Mean Squared Error between the reconstructed data \(\hat{x}\) and original data \( x \). \(\mathcal{L_{\text{KLD}}}\) is the KL divergence between the learned latent distribution and the prior distribution. \(\mathcal{L_{\text{BCE}}}\) is the Binary Cross-Entropy loss for the classification task. During the optimization process, by minimizing $\mathcal{L}$, the model simultaneously optimizes the reconstruction loss, KL divergence, and mutual information.

\section{Experiments}\label{sec:exper}
In this section, we present an experimental evaluation of the proposed solution, conducted using six publicly available benchmark datasets with diverse characteristics. We demonstrate the effectiveness of the proposed method and delve into the contributions of each module within the DTD-VAE to gain a deeper understanding of its operational mechanism. Additionally, we performed several in-depth analyses to address the following research questions (RQs):
\begin{itemize}
\item{
\textbf{RQ1}: How does the performance of DTD-VAE compare to SOTA methods in predicting financial risk?
}
\item{
\textbf{RQ2}: What is the contribution of DTD-VAE's key modules to its performance improvement?
}
\item{
\textbf{RQ3}: How effectively does our solution address the challenges described in Section~\ref{sec:intro} on different datasets? 
}
\item{
\textbf{RQ4}: How do DTD-VAE's hyperparameter settings affect its prediction performance?
}
\end{itemize}

\subsection{Experimental Settings}
\subsubsection{Datasets}
Experiments were conducted on six well-known public benchmark datasets, each showcasing unique characteristics to avoid confidentiality issues. Table \ref{tab:dataset_characteristics} provides a statistical overview of the datasets after pre-processing, where applicable. The preprocessing steps included: \textbf{i)} deleting meaningless features (e.g., loan ID, customer ID) and those with a miss rate above 60\%; \textbf{ii)} imputing missing values using mean/mode replacement for numeric/nominal features; and \textbf{iii)} encoding Boolean features as 0 or 1, standardizing continuous features with min-max normalization, and converting multi-valued and disordered features into one-hot codes. After preprocessing, We considered class labels for these datasets: Lending Club (LC), Prosper Loan (PL), Bank Loan Status (BLS), Bank Fears Loanliness (BFL), Give Me Some Credit (GMSC), and Santander Customer Transaction (SCT). For the Lending dataset (2017Q1), ‘‘LoanStatus'' (seven categories) was treated as the class label. We focused on repaid and defaulted loans, deleting ``current loans'' due to uncertainty about future defaults. ``Fully paid'' loans were defined as ``good credit,’’ and the remaining categories as ``bad credit.'' In the Prosper dataset, ‘‘LoanStatus'' included 12 categories. Similar to Lending, ``cancelled'' and ``current'' loans were deleted. ``Completed'' loans were defined as ``good credit,'' and the remaining categories as ``bad credit.'' For the Bank Loan Status, Bank Fears Loanliness, Give Me Some Credit, and Santander Customer Transaction datasets, loans were already divided into good and bad credit classes, so no additional class label handling was needed. In this study, the good and bad credit samples were labeled as 0 and 1, respectively.

\begin{table}[t]
\scriptsize
\centering
\caption{\footnotesize Statistics of dataset after preprocessing}
\label{tab:dataset_characteristics}
\begin{tabular}{c|cccccc}
\toprule
Dataset & LC & PL & BLS & BFL & GMSC & SCT \\
\midrule
Features & 145 & 81 & 19 & 45 & 12 & 202 \\
Samples  & 96,779 & 113,937 & 100,514 & 532,428 & 150,000 & 200,000 \\
Good     & 23,381 & 38,074 & 77,361 & 406,601 & 139,974 & 179,902 \\
Bad      & 9,088 & 19,282 & 22,639 & 125,827 & 10,026 & 20,098 \\
IR       & 2.57 & 1.97 & 3.42 & 3.23 & 13.96 & 8.95 \\
\bottomrule
\end{tabular}
\end{table}

\subsubsection{Evaluation Protocol and Metrics}
In our evaluation protocol, we adhered to the metrics employed in previous research~\citep{mancisidor2022generating,zhang2024consumer}. An important function of credit scoring models is to rank customers so that high-risk clients can be prioritized. Probability values provide a natural basis for ranking, whereas metrics such as recall and precision are not directly applicable to this purpose~\citep{mancisidor2022generating,thomas2000survey,siddiqi2012credit}. In our evaluation protocol, we adopted the Area Under the Receiver Operating Characteristic Curve (ROC-AUC) and Accuracy Ratio (AR) to assess the overall performance of the proposed solution. The ROC-AUC is a commonly used evaluation metric based on probability values and effectively reflects the model's ranking ability. The AR provides another measure of the model's discriminatory power. Probability values can be directly interpreted as the likelihood of a customer defaulting, making it easier for banks to understand the model's output and to develop strategies accordingly. These metrics evaluate the model’s effectiveness and reliability in real-world applications.

\subsubsection{Comparison Methods}
To demonstrate the effectiveness of the DTD-VAE, we included three classification methods across distinct categories: \textbf{i)} Traditional Classification Approaches: Comprises well-established algorithms like Logistic Regression (LogR), Random Forest (RF), LightGBM, and eXtreme Gradient Boosting (XGB). These methods are widely recognized for their robustness and versatility. \textbf{ii)} Approaches to Disentangled Temporal Dependencies: Includes advanced deep learning models such as LSTM~\citep{liang2020forecasting}, Informers ~\citep{zhou2021informer}, Ladder VAE~\citep{sonderby2016ladder}, VRNN~\citep{he2018variational} and VAEneu ~\citep{koochali2025vaeneu}. These models excel in learning temporal patterns and latent variable dynamics. \textbf{iii)} Approaches Incorporating the Gating Mechanism: Focuses on models like VanillaVAE, VAE with Mixture of Experts (Crocodile~\citep{lin2024disentangled}), GIB~\citep{alesiani2023gated}, GPGVAE ~\citep{yu2024learning}, GaVaMoE ~\citep{tang2025gavamoe}, and MulVAEK ~\citep{li2025novel}, emphasizing the importance of gating mechanisms for generative tasks. By comparing these models, we highlight the effectiveness of the DTD-VAE in capturing complex temporal dependencies and informative latent structures.

\subsubsection{Hyperparameter Settings}
For each dataset, the data were randomly divided into training (80\%), validation (10\%), and testing (10\%) sets. Within the VAE paradigm, MLPs are implemented for the encoding mechanism and an element-wise gating experts mechanism is devised for the decoding process. A pilot investigation assessed the impact of MLPs depth, identifying 2-hidden layer MLPs with topology \{(64), (128, 64)\} for the encoder and a 2-hidden layer element-wise gating experts structure for the decoders as optimal. This selection was based on the observation that deeper networks did not improve performance but increased hyperparameter tuning and computational demands, which is consistent with the empirical findings of~\citep{liang2018variational}. The dimensionality of the latent space was systematically varied from 4 to 128 for the comparative analysis, and the number of expert was varied from 2 to 64. The ReLU activation function was retained for all hidden layers, demonstrating superior efficacy relative to the Tanh function in our evaluations. Given the binary nature of the data, the outputs from the MLP layer were subjected to sigmoid transformation. Hyperparameter search domains were defined as follows: the learning rate was explored within the interval \( 5 \times 10^{-4} \) to \( 1 \times 10^{-1} \) at logarithmic intervals, and the training epoch was fixed at 20 for optimal performance. The batch size was uniformly set at 128 across all methods and optimized using the Adam optimizer.

\begin{table*}[t]
\scriptsize
\centering
\caption{\footnotesize Our model outperforms SOTA baselines across ROC-AUC(\%) and AR(\%), with improvements in bold and baseline peaks underlined. \textbf{$^*$} indicates that the improvements are statistically significant for $p<0.05$ judged with the runner-up result in each case by paired t-test.}
\label{tab:model_comparison}
\fontsize{10}{16}\selectfont
\resizebox{\textwidth}{!}{%
\begin{tabular}{l|cccccccccccc}
\toprule
\multirow{2}{*}{\textbf{Model}} & \multicolumn{2}{c}{LC} & \multicolumn{2}{c}{PL} & \multicolumn{2}{c}{BLS} & \multicolumn{2}{c}{BFL} & \multicolumn{2}{c}{GMSC} & \multicolumn{2}{c}{SCT} \\
\cmidrule(lr){2-3} \cmidrule(lr){4-5} \cmidrule(lr){6-7} \cmidrule(lr){8-9} \cmidrule(lr){10-11} \cmidrule(lr){12-13}
& ROC-AUC & AR & ROC-AUC & AR & ROC-AUC & AR & ROC-AUC & AR & ROC-AUC & AR & ROC-AUC & AR \\
\midrule
LogR & 63.01±0.92 & 26.02±1.85 & 59.39±0.62 & 18.79±1.25 & 60.51±0.17 & 21.02±0.34 & 59.93±0.77 & 19.87±1.54 & 50.38±0.12 & 0.77±0.25 & 63.57±0.78 & 27.15±1.57 \\
RF & 62.75±0.12 & 25.50±0.25 & 61.00±0.67 & 22.01±1.35 & 61.69±0.32 & 23.38±0.65 & 65.67±0.47 & 31.35±0.95 & 54.34±0.66 & 8.69±1.33 & 65.54±1.13 & 30.89±2.26 \\
LightGBM & 65.23±0.91 & 30.47±1.82 & 62.38±0.63 & 24.76±1.27 & 61.59±0.25 & 23.18±0.50 & 72.98±0.40 & 45.97±0.80 & 54.33±0.78 & 8.67±1.57 & 53.82±1.13 & 7.64±2.27 \\
XGB & 65.37±0.75 & 30.75±1.50 & 62.59±0.49 & 25.19±0.98 & 62.65±0.19 & 25.31±0.39 & 72.19±1.05 & 44.38±2.10 & 54.59±0.55 & 9.19±1.11 & 55.07±1.44 & 10.15±2.89 \\
\midrule
LSTM & 63.90±0.27 & 27.80±0.55 & 61.62±0.48 & 23.24±0.97 & 61.01±0.10 & 22.02±0.20 & 64.69±0.55 & 29.38±1.11 & 52.76±0.39 & 5.53±0.78 & 62.01±0.42 & 24.02±0.85 \\
Informer & 63.44±0.19 & 26.89±0.37 & 61.61±0.34 & 23.22±0.67 & 60.92±0.20 & 21.84±0.40 & 63.08±0.55 & 26.16±1.10 & 52.40±0.57 & 4.80±1.14 & 63.16±0.18 & 26.32±0.35 \\
VanillaVAE & 66.36±0.38 & 32.72±0.73 & 66.85±0.44 & 33.70±0.89 & 72.17±0.48 & 44.34±0.96 & 71.42±0.39 & 42.85±0.79 & 64.05±1.08 & 28.10±2.16 & 76.23±1.11 & 52.47±2.22 \\
LadderVAE & 68.08±0.86 & 36.16±1.72 & 67.09±0.40 & 34.17±0.80 & 72.71±0.25 & 45.43±0.51 & 73.03±0.37 & 46.06±0.73 & 71.14±1.09 & 42.29±2.17 & 77.02±0.87 & 54.05±1.75 \\
VRNN & 67.91±0.19 & 35.82±0.38 & 67.78±0.45 & 35.55±0.90 & 72.31±0.31 & 44.63±0.62 & 73.25±0.23 & 46.51±0.47 & 72.28±0.21 & 44.56±0.42 & 71.90±0.36 & 43.80±0.73 \\
VAEneu & 68.25±0.77 & 36.50±1.54 & 64.93±0.38 & 29.86±0.77 & 68.57±0.63 & 37.14±1.27 & 72.98±0.92 & 45.96±1.85 & 73.65±0.64 & 47.30±1.29 & 77.41±0.42 & 54.82±0.84 \\
\midrule
Crocodile & 66.89±1.13 & 33.78±2.13 & 66.83±0.18 & 33.66±0.37 & 72.08±0.21 & 44.17±0.43 & 71.05±0.32 & 42.10±0.64 & 64.16±1.13 & 28.33±2.26 & 76.47±1.43 & 52.94±2.86 \\
GPGVAE & 66.33±0.54 & 32.66±1.08 & 68.17±0.59 & 36.34±1.19 & \underline{72.21±0.70} & \underline{44.42±1.40} & 69.36±0.66 & 38.72±1.32 & 65.06±1.21 & 30.13±2.41 & 77.33±0.46 & 54.66±0.93 \\
GIB & 68.98±0.48 & 37.97±0.96 & \underline{68.24±0.27} & \underline{36.49±0.54} & 72.68±0.28 & 45.36±0.56 & \underline{75.16±0.39} & \underline{50.32±0.79} & \underline{75.02±0.81} & \underline{50.05±1.63} & \underline{78.93±0.34} & \underline{57.87±0.68} \\
GaVaMoE & 69.07±1.26 & 38.15±2.52 & 67.34±0.17 & 34.69±0.35 & 71.06±0.49 & 42.12±0.98 & 73.14±0.34 & 46.29±0.68 & 72.81±1.30 & 45.62±2.60 & 77.55±0.70 & 55.11±1.41 \\
MulVAEK & \underline{69.61±0.86} & \underline{39.23±1.72} & 67.91±0.55 & 35.82±1.10 & 71.94±0.58 & 43.89±1.16 & 74.32±0.15 & 48.64±0.30 & 73.45±0.95 & 46.90±1.90 & 76.89±0.49 & 53.79±0.98 \\
\midrule
\textbf{DTD-VAE} & \textbf{74.47±0.22$^*$} & \textbf{48.94±0.45$^*$} & \textbf{72.67±0.12$^*$} & \textbf{45.34±0.25$^*$} & \textbf{76.41±0.69$^*$} & \textbf{52.82±1.38$^*$} & \textbf{78.36±0.18$^*$} & \textbf{56.73±0.37$^*$} & \textbf{78.66±0.21$^*$} & \textbf{57.33±0.42$^*$} & \textbf{82.23±0.21$^*$} & \textbf{64.47±0.43$^*$} \\
\bottomrule
\end{tabular}
}
\end{table*}

\subsection{Performance Comparisons (RQ1)}
We compared our approach with several SOTA methods and the results are listed in Table~\ref{tab:model_comparison}. Our DTD-VAE model demonstrates robust financial risk prediction performance, significantly exceeding that of all other methods. This evaluation yields several key insights:

\textit{Performance Advantage of DTD-VAE}: Across all benchmark datasets evaluated, it demonstrates significant advantages in key metrics such as ROC-AUC and AR. In particular, it achieves performance improvements ranging from 3.2\% to 4.86\% in ROC-AUC and 6.41\% to 9.71\% in AR, highlighting its superior performance. It shows competitive performance compared to the second best method, demonstrating its strong potential in real-world financial risk prediction tasks. 
    
\textit{Comparison with Other Models}: Initially, an analysis of the experimental performance of the first group of methods, which focus exclusively on the characteristics of the sample without considering additional information, revealed significant limitations in their efficacy. In contrast, gating-based mechanism methods leverage the combined power of inference and generative models to capture user information, especially latent factors, thereby substantially enhancing prediction precision. Models such as GPGVAE have shown commendable performance on specific datasets. Similarly, the Ladder VAE and VRNN models, with their superior ability to manage temporal dependencies, outperformed the methods in the first and second groups on certain datasets. However, these models face challenges such as disentangling data complexity and temporal dependencies or the difficulty in extracting task-relevant information. By comparison, the incorporation of the ATD and EG mechanisms, as illustrated in Figure ~\ref{fig:_ds_vae_framework_}, allows DTD-VAE to overcome these obstacles, leading to enhanced diversity and improved results.
    
\textit{Performance Analysis Across Datasets}: To evaluate the efficacy of DTD-VAE, we performed experiments across datasets of varying scales, with a particular focus on the small-scale BLS and GMSC datasets (with fewer than 20 features), which have notably few features. In particular, within the BLS dataset, DTD-VAE demonstrated significant gains in ROC-AUC and AR, with increases of 3.7\% and 7.39\%, respectively. Similarly, on the GMSC dataset, DTD-VAE recorded substantial improvements, with AUC and AR improving by 3.64\% and 7.28\%, respectively, highlighting its proficiency in tackling specific dataset challenges. In other datasets, particularly the SCT data set (with more than 200 features), DTD-VAE performed well, with average improvements of 3.3\% in ROC-AUC and 6.6\% in AR. By comparing these results across datasets, we further validated the exceptional performance of the ATD and EG modules, confirming their critical contribution to the superior performance of the DTD-VAE across a diverse array of datasets.
    
\textit{Potential Application Scenarios}: Given DTD-VAE's excellent performance on multiple datasets, it can be widely applied in various financial prediction scenarios, including but not limited to personal loans, personal credit, and personal transactions.

In summary, DTD-VAE not only achieves significant performance improvements technologically but also demonstrates strong prediction capabilities in practical applications, with broad application prospects.

\subsection{Discussion of Model Variants (RQ2)}
\begin{table}[t]
\centering
\scriptsize
\caption{\footnotesize Performance comparison of ablation study}
\label{tab:ablation_study}
\begin{tabular}{lc|ccc|c|ccc}
\toprule
\multirow{2}{*}{Dataset} & \multirow{2}{*}{Metric} & \multicolumn{3}{c}{w/o} & \multirow{2}{*}{DS-VAE} & \multicolumn{3}{c}{Gain} \\
\cmidrule(lr){3-5} \cmidrule(lr){7-9}
 & & ATD & EG & ATD\&EG & & ATD & EG & ATD\&EG \\
\midrule
\multirow{2}{*}{LC} & ROC-AUC & 72.68 & 73.92 & 66.36 & 74.47 & -1.79 & -0.55 & -8.11 \\
   & Gini & 45.36 & 47.85 & 32.72 & 48.94 & -3.58 & -1.09 & -16.22 \\
\multirow{2}{*}{PL} & ROC-AUC & 71.39 & 70.57 & 66.85 & 72.67 & -1.28 & -2.10 & -5.82 \\
   & Gini & 42.78 & 41.14 & 33.70 & 45.34 & -2.56 & -4.20 & -11.64 \\
\multirow{2}{*}{BLS} & ROC-AUC & 74.17 & 75.61 & 72.17 & 76.41 & -2.24 & -0.80 & -4.24 \\
   & Gini & 48.34 & 51.22 & 44.34 & 52.82 & -4.48 & -1.60 & -8.48 \\
\multirow{2}{*}{BFL} & ROC-AUC & 76.30 & 77.59 & 71.42 & 78.36 & -2.06 & -0.77 & -6.94 \\
   & Gini & 52.61 & 55.18 & 42.85 & 56.73 & -4.12 & -1.55 & -13.88 \\
\multirow{2}{*}{GMSC} & ROC-AUC & 78.38 & 77.31 & 64.05 & 78.66 & -0.28 & -1.35 & -14.61 \\
   & Gini & 56.76 & 54.62 & 28.10 & 57.33 & -0.57 & -2.71 & -29.23 \\
\multirow{2}{*}{STC} & ROC-AUC & 80.22 & 81.04 & 76.23 & 82.23 & -2.01 & -1.19 & -6.00 \\
   & Gini & 60.45 & 62.09 & 52.47 & 64.47 & -4.02 & -2.38 & -12.00 \\
\bottomrule
\end{tabular}
\vspace{-1.5em}
\end{table}

We conducted an ablation study on different submodules in the DTD-VAE to provide a detailed analysis of its function and efficiency. The variant models of the DTD-VAE consist of the following structures, and the notation is simply for simplicity:
\begin{itemize}
    \item \textbf{DTD-VAE (w/o ATD)}: removing ATD Module, as described Section ~\ref{sec:intro}, aims to enhance the model’s understanding of the data’s intrinsic structure. 
    \item \textbf{DTD-VAE (w/o EG)}: removing EG Module module and replacing with a standard decoder.
    \item \textbf{DTD-VAE (w/o ATD\&EG)}: removing the ATD Module and EG Module.
    \item \textbf{DTD-VAE}: complete structure of DTD-VAE.
\end{itemize}

As shown in Table~\ref{tab:ablation_study}, the complete structure of the DTD-VAE outperformed all other DTD-VAE variants, and we can draw the following conclusions for each submodule:

\textit{Impact of ATD on Performance} ($Gain_{ATD}$): The data consistently shows that incorporating ATD mechanism yields significant positive gains across various datasets and evaluation metrics, such as ROC-AUC and AR. Specifically, the average decreases observed are 1.13\% and 2.26\%, respectively. This consistent enhancement suggests that the ATD mechanism is crucial for effectively capturing temporal dependencies among latent variables, thereby deepening the model’s comprehension of the underlying data structure. The persistent positive impact of ATD on these metrics highlights its indispensable role in boosting the precision of the predictions.

\textit{Impact of EG on Performance}($Gain_{EG}$): The EG mechanism demonstrates a substantial ability to enhance the model's generative capabilities. The gains observed when the EG is integrated are consistently positive, indicating that the mechanism significantly contributes to improved performance across various metrics. Specifically, the average decreases of 1.61\% in ROC-AUC and 3.22\% in AR. This suggests that the EG mechanism, through its element-wise gating experts approach, significantly amplifies the model's capacity to generate high-quality predictions that closely align with the user preferences.

\textit{Combined Impact of ATD and EG }($Gain_\text{ATD\&EG}$): The combined impact of dual mechanism is particularly noteworthy, often resulting in the most substantial gains or minimal losses across various datasets. This highlights the high efficacy of their collaboration in optimizing the performance of DTD-VAE, particularly in metrics such as ROC-AUC and AR, with average decreases of 7.62\% and 15.24\%, respectively. This underscores the significant positive impact of both mechanisms working together to enhance the model's predictive capabilities.

In summary, the analysis of $Gain_{ATD}$, $Gain_{\text{EG}}$, and $Gain_\text{ATD\&EG}$ highlights the critical roles of ATD and EG in the performance of the DTD-VAE. The ATD's ability to capture the temporal dependency among latent variables enhances the model's representation learning, whereas EG improves its generative capabilities. The combined effect of these mechanisms further optimizes the performance of the DTD-VAE, making it a powerful tool for financial risk prediction.

\subsection{Discussion of In-Depth Studies (RQ3)}
\begin{table*}[t]
\centering
\scriptsize
\centering
\caption{\footnotesize Comprehensive evaluation of traditional methods, augmented by DTD-VAE-generated insights, in tackling the financial risk prediction challenge.}
\label{tab:traditional_model_dtd_vae}
\fontsize{10}{16}\selectfont
\resizebox{\textwidth}{!}{%
\begin{tabular}{l|ccccccccccccc}
\toprule
\multirow{2}{*}{\textbf{Model}} & \multicolumn{2}{c}{LC} & \multicolumn{2}{c}{PL} & \multicolumn{2}{c}{BLS} & \multicolumn{2}{c}{BFL} & \multicolumn{2}{c}{GMSC} & \multicolumn{2}{c}{SCT} \\
\cmidrule(lr){2-3} \cmidrule(lr){4-5} \cmidrule(lr){6-7} \cmidrule(lr){8-9} \cmidrule(lr){10-11} \cmidrule(lr){12-13}
& ROC-AUC & AR & ROC-AUC & AR & ROC-AUC & AR & ROC-AUC & AR & ROC-AUC & AR & ROC-AUC & AR \\
\midrule
LogR & 63.01±0.92 & 26.02±1.85 & 59.39±0.62 & 18.79±1.25 & 60.51±0.17 & 21.02±0.34 & 59.93±0.77 & 19.87±1.54 & 50.38±0.12 & 0.77±0.25 & 63.57±0.78 & 27.15±1.57 \\
RF & 62.75±0.12 & 25.50±0.25 & 61.00±0.67 & 22.01±1.35 & 61.69±0.32 & 23.38±0.65 & 65.67±0.47 & 31.35±0.95 & 54.34±0.66 & 8.69±1.33 & 65.54±1.13 & 30.89±2.26 \\
LightGBM & 65.23±0.91 & 30.47±1.82 & 62.38±0.63 & 24.76±1.27 & 61.59±0.25 & 23.18±0.50 & 72.98±0.40 & 45.97±0.80 & 54.33±0.78 & 8.67±1.57 & 53.82±1.13 & 7.64±2.27 \\
XGB & 65.37±0.75 & 30.75±1.50 & 62.59±0.49 & 25.19±0.98 & 62.65±0.19 & 25.31±0.39 & 72.19±1.05 & 44.38±2.10 & 54.59±0.55 & 9.19±1.11 & 55.07±1.44 & 10.15±2.89 \\
\midrule
LogR(DTD-VAE) & 66.96±0.78 & 33.92±1.57 & 63.58±0.39 & 27.16±0.78 & 62.36±0.82 & 24.72±1.64 & 65.30±0.71 & 30.60±1.42 & 54.93±3.13 & 9.86±6.27 & 64.29±0.38 & 28.57±0.76 \\
RF(DTD-VAE) & 67.62±1.56 & 35.25±3.13 & 64.10±0.32 & 28.21±0.63 & 64.91±0.23 & 29.81±0.46 & 67.15±0.35 & 34.31±0.69 & 54.99±0.41 & 9.98±0.82 & 68.04±0.35 & 36.07±0.69 \\
LightGBM(DTD-VAE) & 65.64±0.23 & 31.27±0.46 & 63.06±0.42 & 26.12±0.83 & 62.42±0.20 & 24.84±0.41 & 73.05±0.27 & 46.09±0.53 & 54.36±0.13 & 8.72±0.26 & 54.54±0.24 & 9.09±0.48 \\
XGB(DTD-VAE) & 65.77±0.37 & 31.53±0.74 & 63.16±0.22 & 25.27±0.53 & 62.90±0.24 & 25.81±0.49 & 73.48±0.30 & 46.95±0.59 & 55.49±0.30 & 10.98±0.60 & 55.52±0.31 & 11.04±0.63 \\
\bottomrule
\end{tabular}}
\end{table*}
\subsubsection{Flexibility of DTD-VAE}
We explored the potential for performance enhancement by integrating the results generated by the DTD-VAE into various traditional models. As illustrated in Table~\ref{tab:traditional_model_dtd_vae}, the traditional models consistently outperformed their counterparts without these enhancements on all metrics. This finding underscores that the seamless integration of DTD-VAE into existing methods yields substantial performance improvements, expanding the applicability of the proposed framework.

\begin{figure}
\centering
\includegraphics[width=0.6\columnwidth]{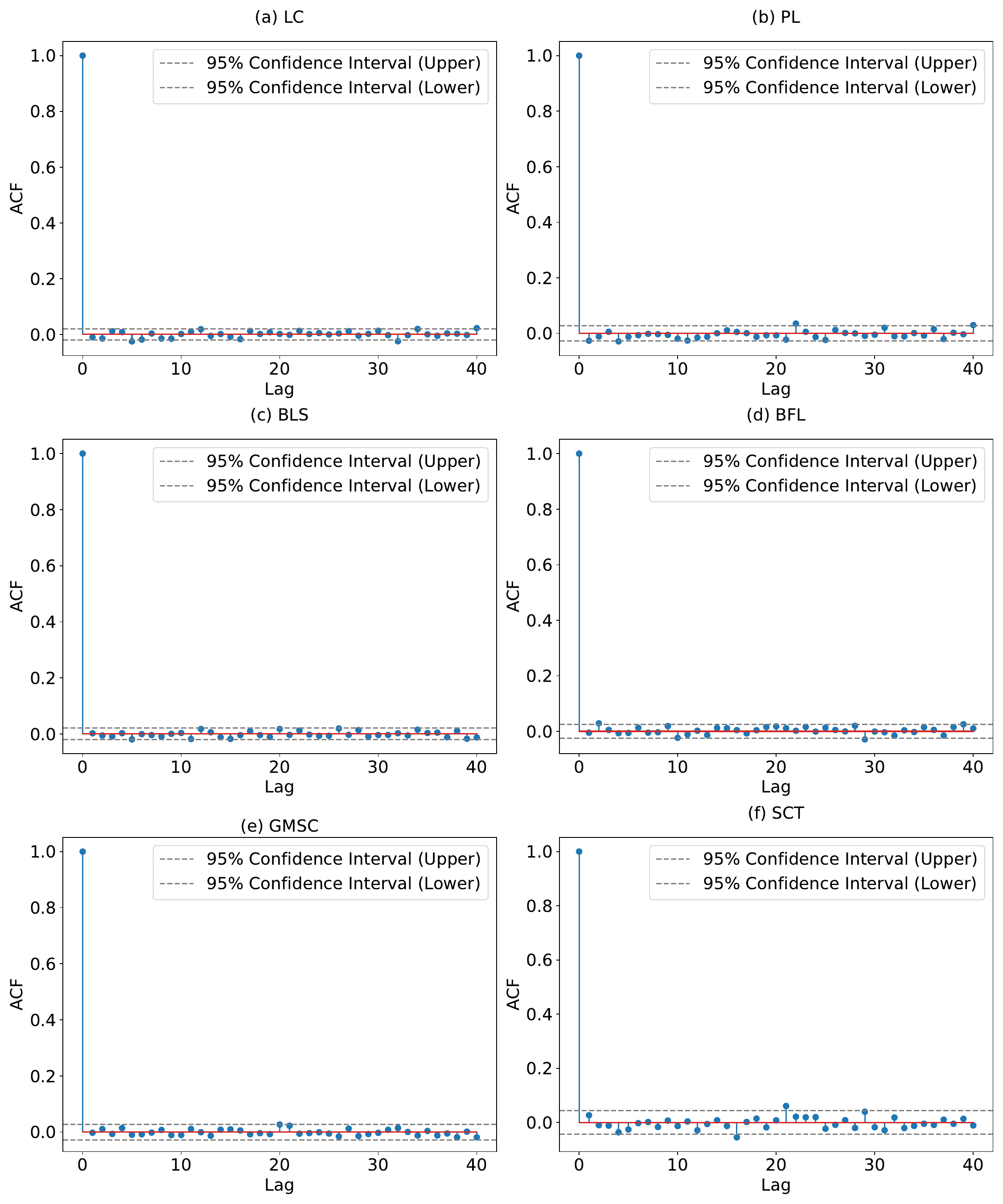}
\caption{\footnotesize ACF analysis of temporal dependencies in ATD module across datasets.}
\label{fig:_atd_vis_acf_}
\end{figure}

\begin{figure*}
\centering
\includegraphics[width=\linewidth]{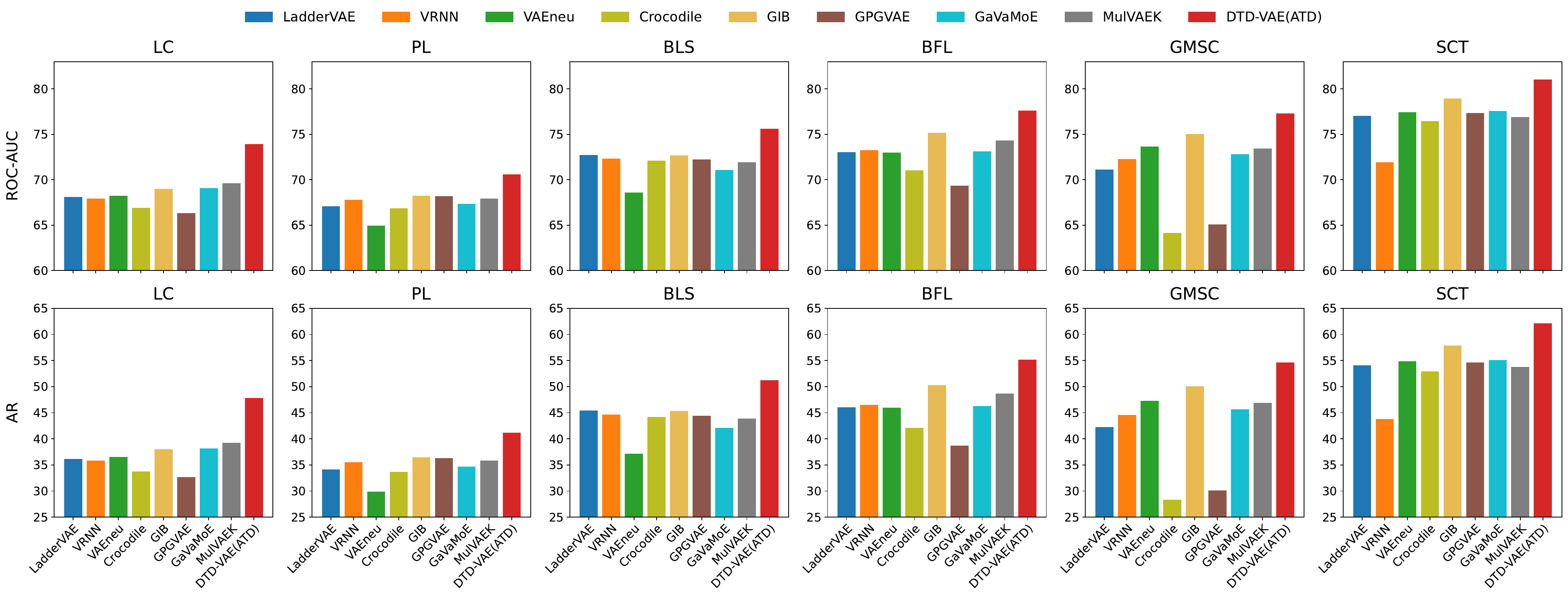}
\caption{\footnotesize Performance comparison of DTD-VAE (Only with ATD) with SOTA VAE-based models across all datasets based on ROC AUC and AR.}
\label{fig:_atd_prediction_vis_}
\end{figure*}

\subsubsection{The Analysis of ATD Mechanism} 
Traditional machine learning methods (e.g., RF and LightGBM) struggle with data complexity and temporal dependencies in loan-default prediction. They often neglect sequential patterns in historical data, including credit scores and repayment records, and thus fail to capture how past behaviors influence current risk. This limitation leads to a degradation in predictive accuracy over time. To address this issue, this study introduces the ATD mechanism, and the experimental results demonstrate its ability to model temporal dynamics. The capability of the ATD mechanism to capture temporal dependencies was validated using autocorrelation function (ACF) analysis. As shown in Fig.~\ref{fig:_atd_vis_acf_}, the ACF values decay rapidly after lag 1, indicating a short-term dependency consistent with an AR(1) process, with a significant temporal correlation at lag 1 for the time-series variables. Furthermore, by leveraging past latent variables to predict future variables, the ATD enhances the predictive accuracy in the latent space. As demonstrated in Fig.~\ref{fig:_atd_prediction_vis_}, the integration of ATD into DTD-VAE effectively captures temporal dynamics and complex data structures, yielding superior credit risk prediction performance compared to existing models, including Ladder VAE~\citep{sonderby2016ladder}, VRNN~\citep{he2018variational}, VAneu~\citep{koochali2025vaeneu}, Crocodile~\citep{lin2024disentangled}, GIB~\citep{alesiani2023gated}, GPGVAE~\citep{yu2024learning}, GaVaMoE~\citep{tang2025gavamoe}, and MulVAEK~\citep{li2025novel}.

\begin{figure}
\centering
\includegraphics[width=0.6\columnwidth]{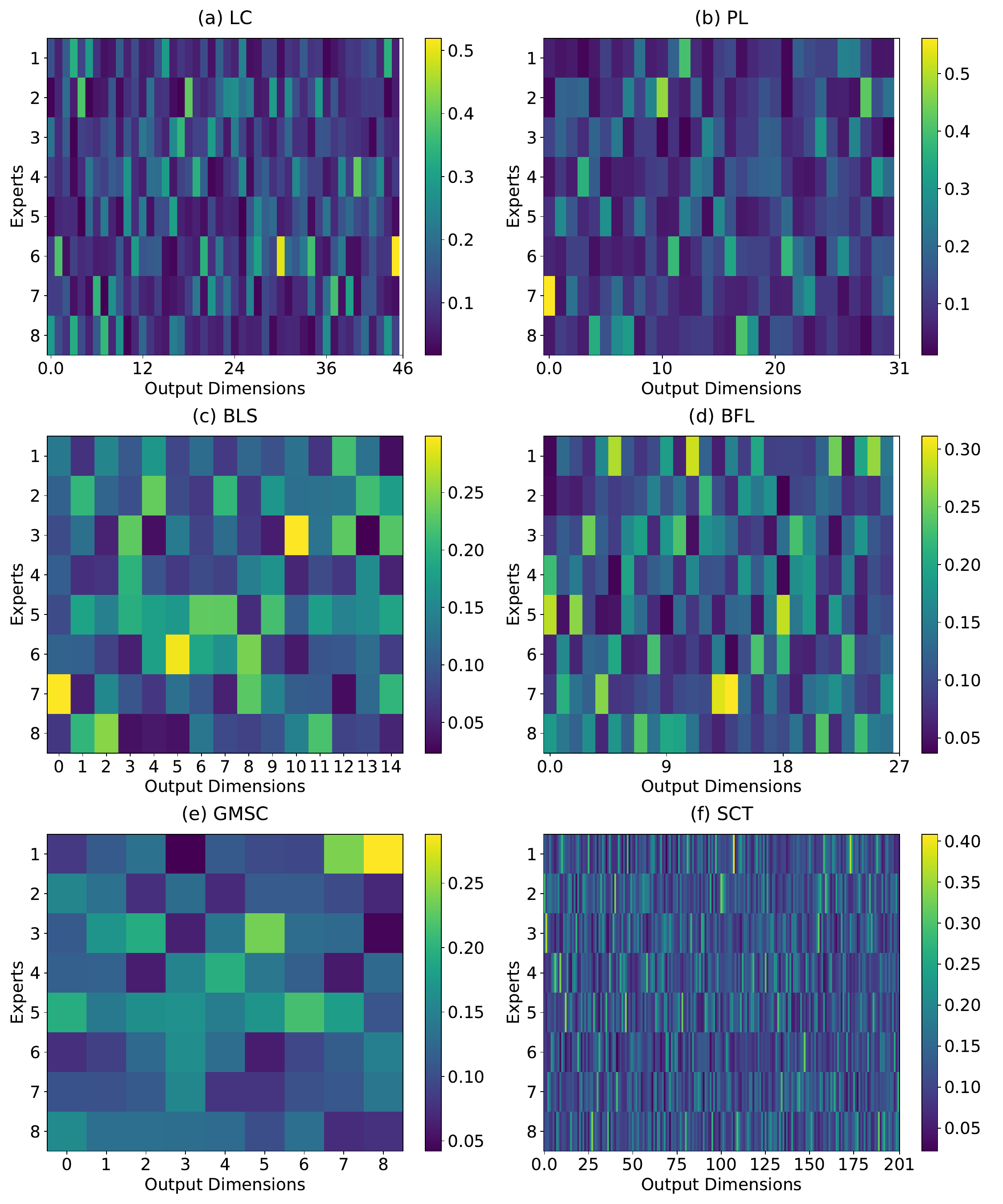}
\caption{\footnotesize Heatmap visualization of EG mechanisms assigning independent weights to each dimension of each expert model across all datasets.}
\label{fig:_eg_gating_exp_vis_}
\end{figure}

\begin{figure*}
\centering
\includegraphics[width=\linewidth]{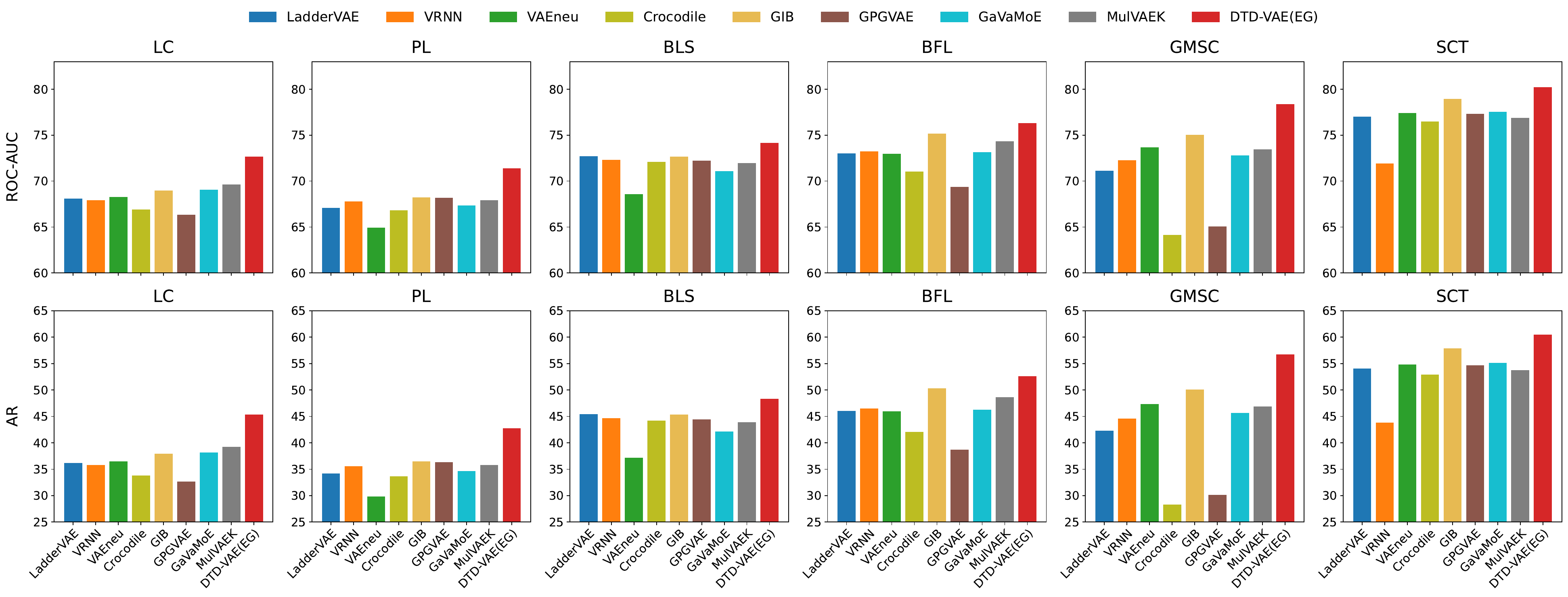}
\caption{\footnotesize Performance comparison of DTD-VAE (Only with EG) with SOTA VAE-based models across all datasets based on ROC AUC and AR.}
\label{fig:_moe_eg_metirc_two_pic_}
\end{figure*}

\begin{figure}[t]
    \centering
    \begin{minipage}[b]{0.45\columnwidth}
        \centering
        \includegraphics[width=0.9\textwidth]{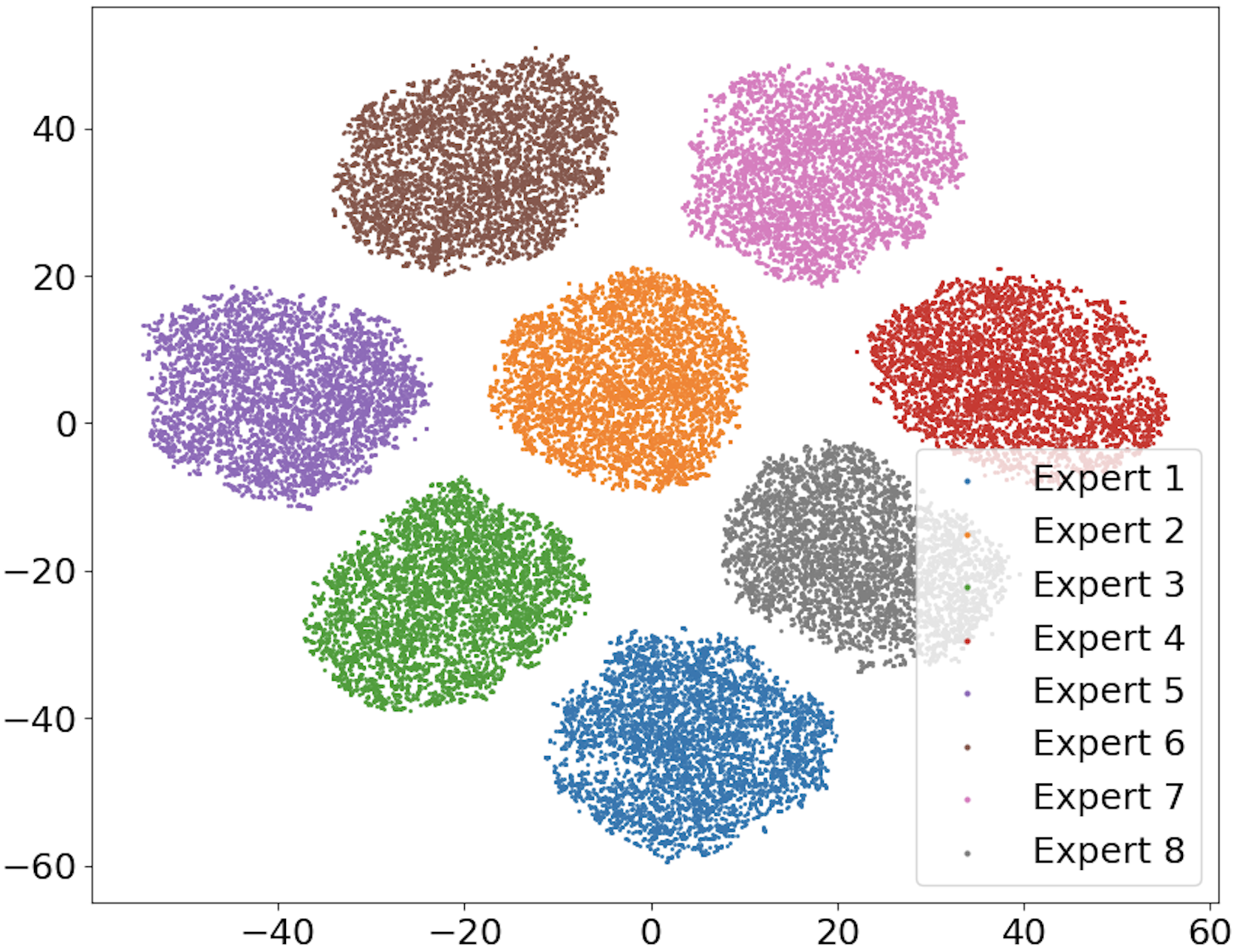}
    \end{minipage}
    \hspace{0.01cm}
    \begin{minipage}[b]{0.45\columnwidth}
        \centering
        \includegraphics[width=0.9\textwidth]{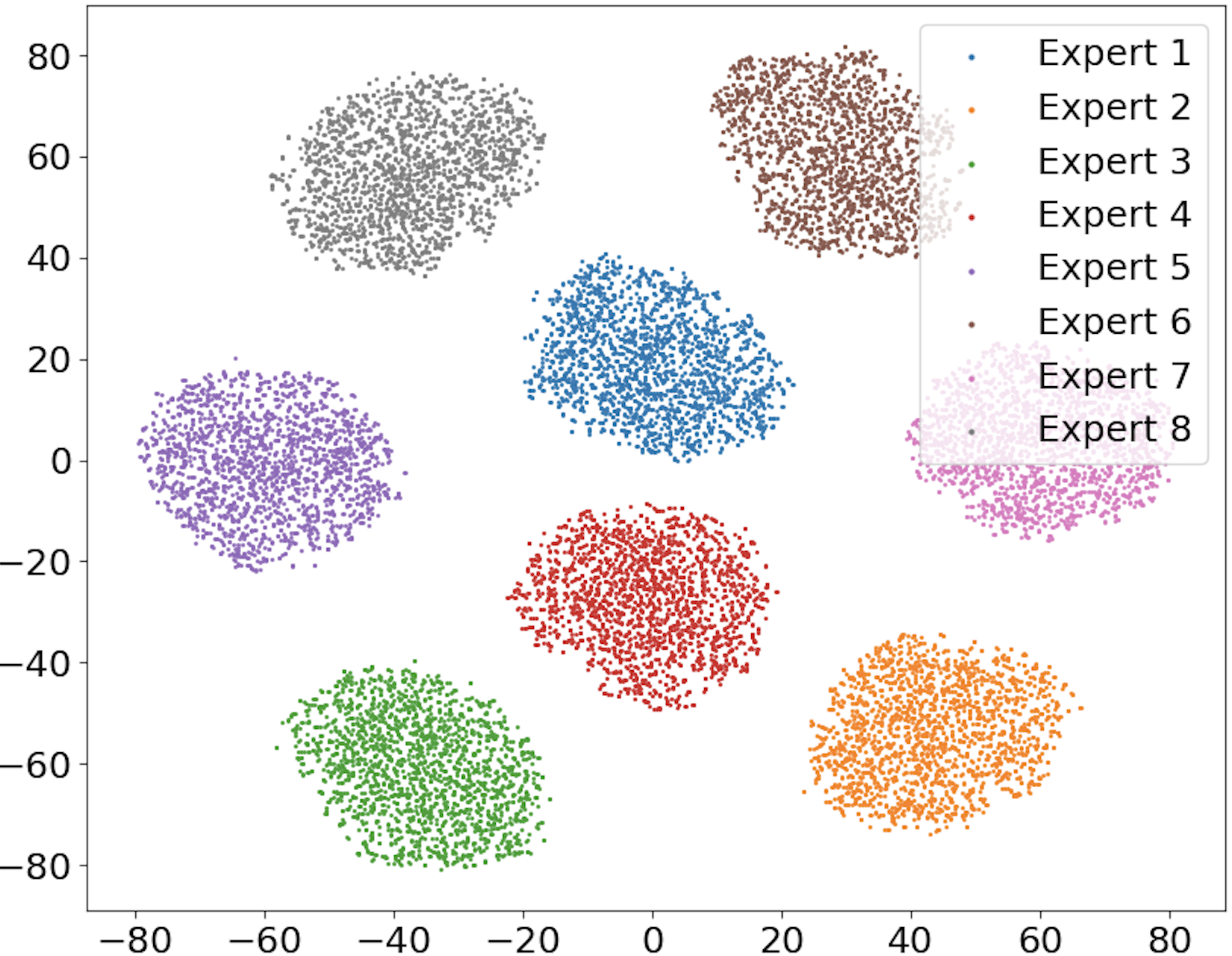}
    \end{minipage}
    \vspace{0.01cm}
    \begin{minipage}[b]{0.45\columnwidth}
        \vspace{-2em}
        \centering
        \includegraphics[width=0.9\textwidth]{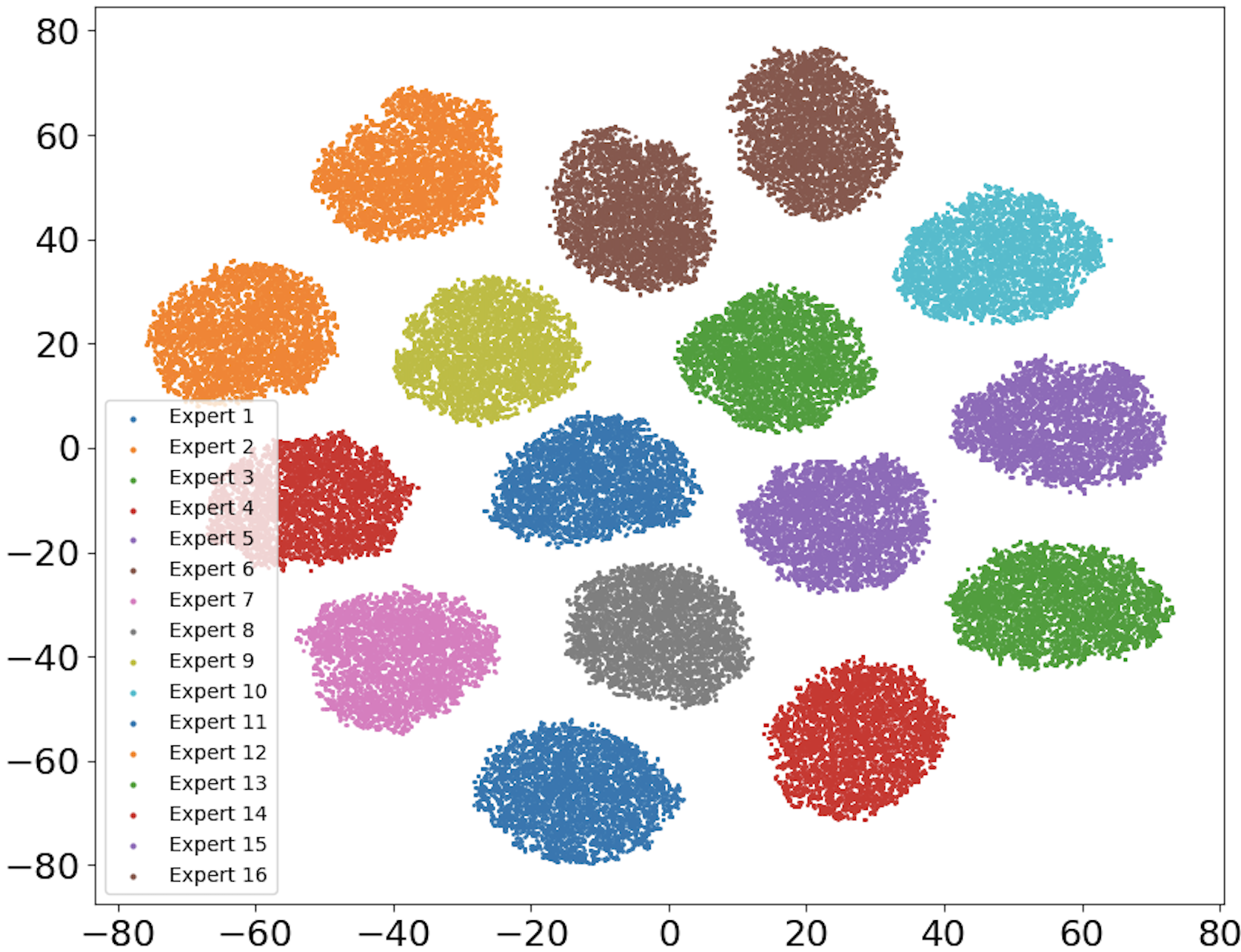}
    \end{minipage}
    \vspace{0.01cm}
    \begin{minipage}[b]{0.45\columnwidth}
        \vspace{1em}
        \centering
        \includegraphics[width=0.9\textwidth]{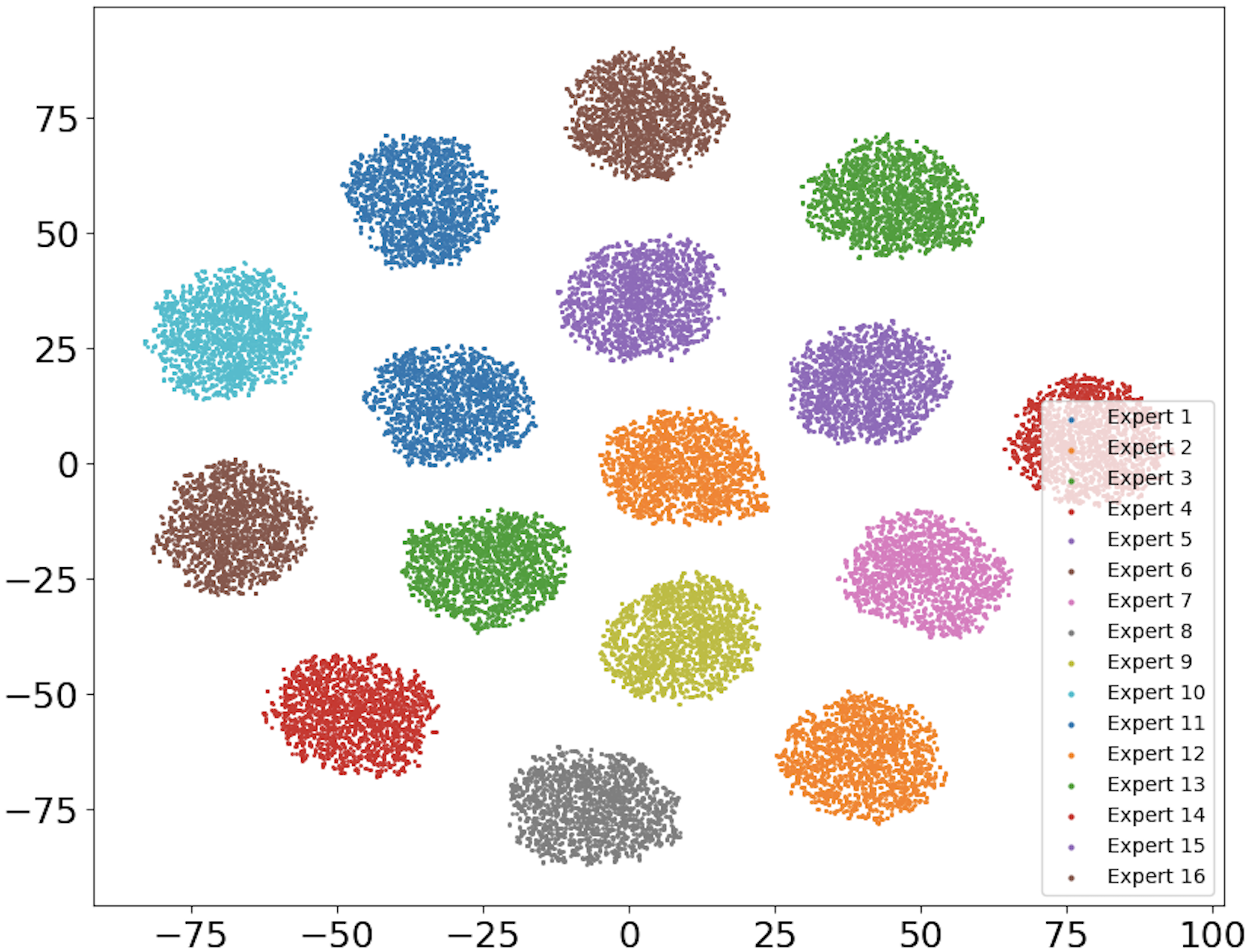}
    \end{minipage}

    \caption{\footnotesize Visualization of new samples generated by 8 and 16 experts in an EG decoder layer on the PL and STC datasets.}
    \label{fig:_eg_expert_vis_}
\end{figure}

\subsubsection{The Analysis of EG Mechanism} 
The EG mechanism enhances model interpretability by providing clear insights into how different experts contribute to capturing variations in user preferences. Unlike traditional gating mechanisms that assign global weights uniformly across all dimensions, EG assigns independent weights to each latent dimension, enabling fine-grained, dimension-level control. This approach allows for more flexible and precise regulation of feature representations, as shown in Fig.~\ref{fig:_eg_gating_exp_vis_}. By supporting dimension-level disentangled representation learning, EG ensures that each latent dimension corresponds to an interpretable data feature, improving both model interpretability and performance. As demonstrated in Fig.~\ref{fig:_eg_expert_vis_}, this leads to more meaningful latent structures. Furthermore, EG achieves competitive performance compared to existing models~\citep{sonderby2016ladder,he2018variational,koochali2025vaeneu,lin2024disentangled,alesiani2023gated,yu2024learning,tang2025gavamoe,li2025novel}.

\begin{figure}[t]
\centering
\includegraphics[width=0.7\columnwidth]{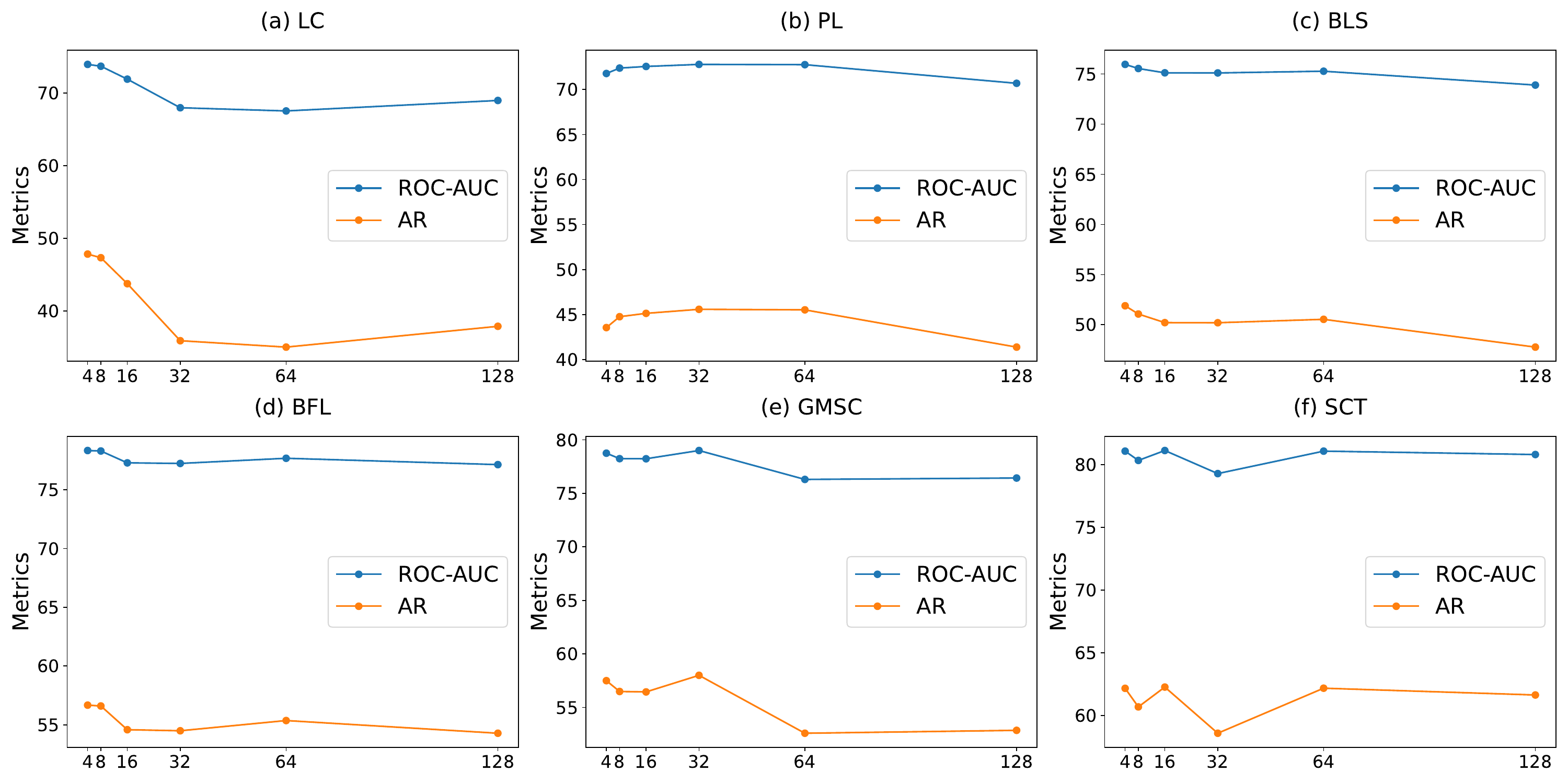}
\caption{\footnotesize Parameter sensitivity analysis across six datasets on latent space size.}
\label{fig:_parameter_sensitivity_latent_}
\end{figure}

\begin{figure}[h]
\centering
\includegraphics[width=0.7\columnwidth]{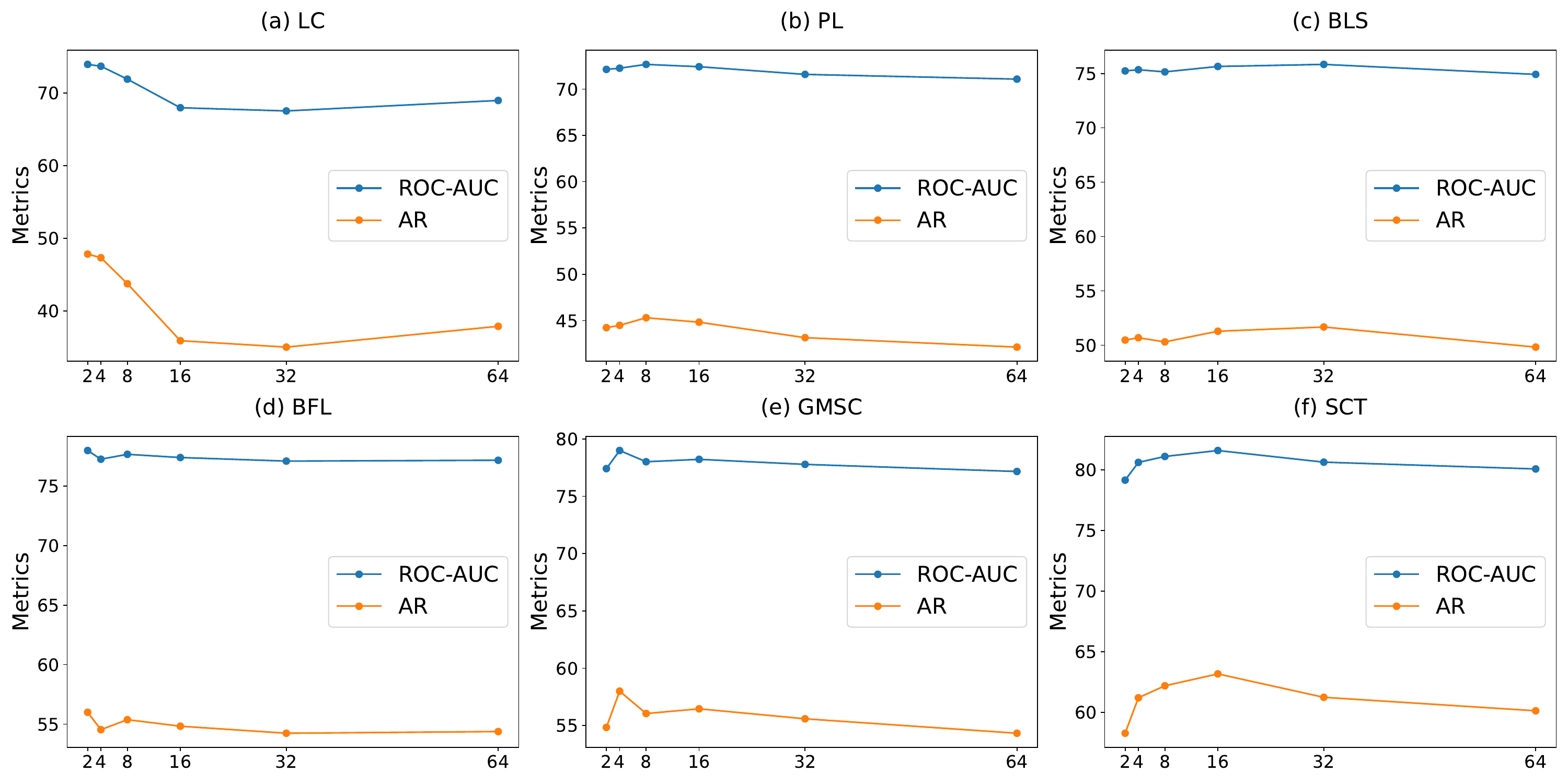}
\caption{\footnotesize Parameter sensitivity analysis across six datasets on expert size.}
\label{fig:_parameter_sensitivity_experts_}
\end{figure}

\begin{figure}[h]
\centering
\includegraphics[width=0.7\columnwidth]{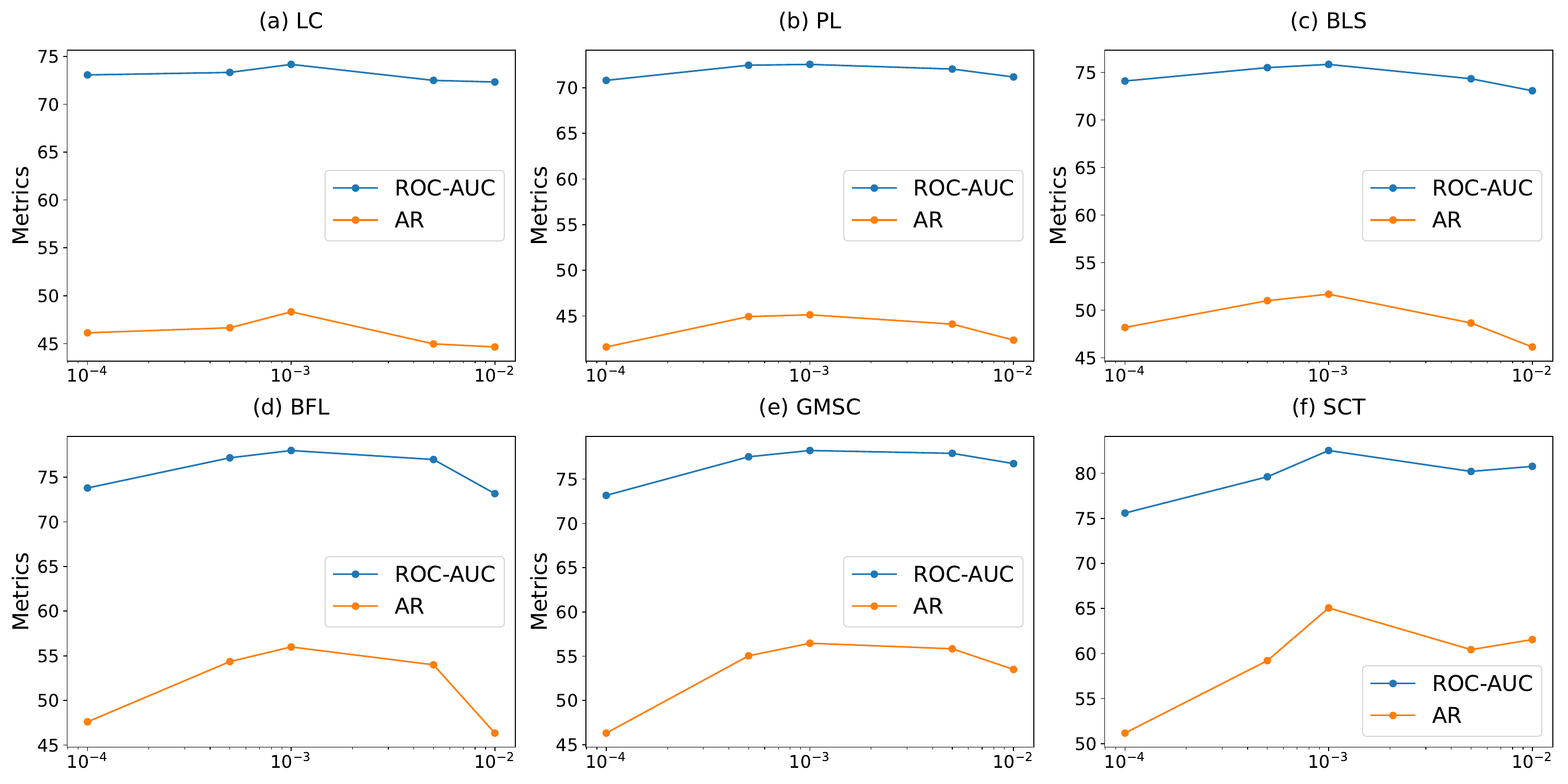}
\caption{\footnotesize Learning rate sensitivity was evaluated across six datasets.}
\label{fig:_parameter_sensitivity_lr_}
\end{figure}

\subsection{Hyperparameter Analysis (RQ4)}
To confirm the correlation between model performance and specific experimental settings, we conducted additional experiments to investigate the sensitivity of the performance to various key hyperparameters.

\textit{Impact of the Latent Space Size}. To ascertain the influence of distinct latent space sizes on the functionality of financial risk prediction for various data groups, we evaluated the critical metrics appropriate for each set, as shown in Figs. \ref{fig:_parameter_sensitivity_latent_}. Optimal latent sizes were identified by tracing the highest metric peaks associated with performance. For the LC, PL, BLS, BFL, GMSC, and STC datasets, the standout results were obtained with latent dimensions of 4, 32, 4, 4, 4, and 16, respectively. These outcomes highlight the necessity of precise latent dimension adjustments tailored uniquely to each dataset as a pivotal factor in amplifying the proficiency of model performance.
    
\textit{Impact of the Number of Experts Size}. In addition, to assess the influence of the number of experts on the prediction performance across diverse datasets, we scrutinized the pivotal metrics for each dataset, as shown in Figs. \ref{fig:_parameter_sensitivity_experts_}. The optimal number of experts was determined by identifying the performance zenith in these metrics. LC, PL, BLS, BFL, GMSC, and STC exhibited peak performances with 2, 8, 32, 2, 4, and 16 experts, respectively, underscoring the dataset-specific optimal configuration of experts that maximizes prediction efficacy. These observations emphasize the criticality of calibrating the number of experts to the idiosyncrasies of each dataset to achieve the pinnacle of performance.
    
\textit{Impact of the Hyperparameter Learning Rate (lr)}. To explore how different lr values influence the efficacy of the proposed approach on various datasets, we extensively analyzed the pertinent performance indicators, as illustrated in Fig. \ref{fig:_parameter_sensitivity_lr_}. The most suitable beta value for each dataset was determined by identifying the junction at which the efficiency peak was reached. Specifically, the peak performance for each of the LC, PL, BLS, BFL, GMSC, and SCT datasets is achieved when the lr is set to $10^{-3}$. These findings underline the significance of custom-tailoring the lr value to complement the unique characteristics intrinsic to each dataset to ensure the peak implementation of the prediction task.

Table \ref{tab:model_comparison} and the parameter sensitivity analyses show that DTD-VAE consistently outperforms baselines, even with suboptimal hyperparameters. It indicates that the DTD-VAE maintains high performance stability across a wide range of settings. Therefore, DTD-VAE can be considered a robust and reliable solution for a specified task.

\subsection{Discussion}
We discuss the implications for practitioners and researchers.

\textbf{Theoretical Significance.} The DTD-VAE is a significant advancement in variational autoencoder frameworks by integrating ATD module with EG module. This integration effectively addresses the limitations of traditional VAE models in capturing intricate temporal dependencies among latent variables and extracting pertinent information. The ATD module enhances the model's capability to learn the intrinsic structure of data through dynamic temporal patterns, whereas the EG mechanism facilitates finer-grained disentanglement of latent variables by assigning independent weights to each dimension. Collectively, these mechanisms improve the quality and diversity of generated samples, enhancing interpretability and controllability. The theoretical contributions of DTD-VAE opens up new research directions for exploring sophisticated latent variable models capable of handling sequential data and complex feature interactions, which are essential for applications ranging from finance to other applications.

\textbf{Practical Implications.} From a practical perspective, the benefits of DTD-VAE can be summarized as follows:
\begin{itemize}
    \item \textit{Enhanced Risk Prediction Accuracy.} Extensive experiments conducted across six public datasets demonstrate that DTD-VAE surpasses SOTA baselines in terms of ROC-AUC and AR metrics, indicating its superior performance in predicting loan default risks. This enhanced accuracy is crucial for financial institutions seeking to refine their decision-making processes.
    \item \textit{Robustness and Adaptability.} The modular architecture of DTD-VAE allows for seamless integration into existing models, making it adaptable to diverse scales and types of data. This flexibility enables customization based on specific application contexts, increasing its practical value across various financial scenarios.
    \item \textit{Parameter Sensitivity Analysis.} Through rigorous parameter sensitivity analysis, we have identified optimal configurations for latent space dimensions and network depth, ensuring stable and reliable performance across different hyperparameter settings. This optimization process is critical for maintaining efficiency and stability in real-world applications.
    \item \textit{Broad Application Scenarios.} Beyond credit risk assessment, DTD-VAE exhibits promise in other areas where temporal dependencies and informative latent structures are pivotal, such as healthcare analytics, customer behavior prediction, and market trend forecasting. Its broad applicability underscores the potential for further research and development in these domains.
\end{itemize}
Overall, DTD-VAE not only provides new perspectives and methods for the theoretical framework of financial risk prediction but also demonstrates its exceptional performance and broad applicability in practical applications. By introducing innovative mechanisms and technologies, the model's effectiveness and practicality are further enhanced.

\section{Conclusion and Future Work}\label{sec:conclusion}
This study proposes the DTD-VAE, an end-to-end framework for credit risk prediction that effectively captures temporal dependencies and disentangles task-relevant features from customer data. By integrating the ATD module and EG mechanism, the model achieves fine-grained latent disentanglement and superior predictive performance. Extensive experiments demonstrate that DTD-VAE outperforms SOTA baselines on benchmark datasets, with ablation studies, efficiency evaluations, and module migration tests validating the contribution and adaptability of its components. Future work will focus on enhancing the model for heterogeneous financial data, improving generative diversity, and increasing interpretability for real-world applications.

\bibliographystyle{elsarticle-harv}
\bibliography{references}

\end{document}